\documentclass[twocolumn,amsthm]{autart}
\usepackage{cite}
\usepackage{amsmath,amssymb,amsfonts}
\usepackage{graphicx}
\usepackage{textcomp}
\usepackage{xcolor}

\usepackage{subfigure}
\usepackage{float}
\usepackage{overpic}
\usepackage{amsthm}
\usepackage{appendix}
\usepackage{mathtools}
\usepackage{tikz}
\usepackage{pgfplots}   
\usepackage{booktabs} 

\newtheorem{theorem}{Theorem}
\newtheorem{lemma}{Lemma}
\newtheorem{remark}{Remark}
\newtheorem{definition}{Definition}
\newtheorem{corollary}{Corollary}

\pgfplotsset{compat=newest}   
\usetikzlibrary{plotmarks}   
\usetikzlibrary{arrows.meta}   
\usepgfplotslibrary{patchplots} 
\usetikzlibrary{angles,quotes,matrix}

\usetikzlibrary{shapes.geometric}
\pgfdeclareplotmark{mystara}{
    \node[star,star point ratio=2.25,minimum size=6pt,
          inner sep=0pt,fill=red] {};
}
\pgfdeclareplotmark{mystard}{
    \node[star,star point ratio=2.25,minimum size=6pt,
          inner sep=0pt,fill=blue] {};
}
\pgfdeclareplotmark{mystarg}{
    \node[star,star point ratio=2.25,minimum size=6pt,
          inner sep=0pt,fill=gray] {};
}
\pgfdeclareplotmark{mystargre}{
    \node[star,star point ratio=2.25,minimum size=6pt,
          inner sep=0pt,fill=green] {};
}
\pgfdeclareplotmark{mystarp}{
    \node[star,star point ratio=2.25,minimum size=6pt,
          inner sep=0pt,fill=purple] {};
}

\begin{document}
\begin{frontmatter}
\title{Reach-avoid games for players with damped double integrator dynamics\thanksref{footnoteinfo}}

\thanks[footnoteinfo]{This work is supported in part by the National Natural Science Foundation of China under Grant U24B20173 and 62573016.
 (Corresponding author: Zongying Shi.)}

\author[Tsinghua]{Mengxin Lyu}\ead{lvmx23@mails.tsinghua.edu.cn},
\author[Tsinghua]{Ruiliang Deng}\ead{drl20@mails.tsinghua.edu.cn},
\author[Tsinghua]{Zongying Shi}\ead{szy@mail.tsinghua.edu.cn},
\author[Tsinghua]{Yisheng Zhong}\ead{zys-dau@mail.tsinghua.edu.cn}

\address[Tsinghua]{Department of Automation, Tsinghua University, Beijing 100084}
\maketitle

\begin{abstract}
This paper investigates a reach–avoid game between two players with damped double integrator dynamics. An optimal state-feedback strategy is derived using a differential game framework combined with geometric analysis. To facilitate the analysis, we introduce the concept of multiple reachable region by characterizing the motion of players with damped double integrator dynamics. Based on this, a new type of the attacker’s dominance region is introduced. We show that distinct strategies are required depending on the location of the terminal position within different areas of the attacker’s dominance region. Furthermore, we prove that the proposed strategies satisfy the necessary condition for optimality. Numerical simulations are provided to illustrate the conclusions.
\end{abstract}

\end{frontmatter}
% \begin{IEEEkeywords}
% reach-avoid games, differential games, optimal control, multi-agent systems
% \end{IEEEkeywords}

\section{Introduction}
Differential games are an important topic with extensive applications, including collision avoidance \cite{ex1}, active defense \cite{ex2}, and region surveillance \cite{ex3}. One type of differential game is the reach-avoid game, in which intelligent agents are divided into defenders and attackers. The defenders aim to protect a target region from the attackers, while the attackers attempt to reach the target before being intercepted by the defenders. 

Reach-avoid differential games were originally formulated in \cite{Isaacs}, where the Hamilton-Jacobi-Isaacs (HJI) equation was proposed to derive the value function and equilibrium strategies. 
For players with simple motion models, geometric methods can be employed to provide intuitive solutions \cite{Dorothy,Shishika2} and to solve games with different target shapes \cite{Shishika,Yan4,Lee,Wang,Dave,Yan2024Obstacles}. Scenarios involving multiple defenders or attackers have also been explored using geometric methods and HJI equations \cite{Yan, Garcia}. For instance, Deng et al. \cite{Deng} proved the optimal strategy for a game involving two attackers and one defender and provided a detailed discussion of the singular surfaces by combining geometric methods with the viscosity solution method. In more general multiplayer settings, the problem can be decomposed into several sub-problems, each involving one attacker, and solved via matching algorithms \cite{MoChen,Yan2,MoChen2,Yan3}. 

Although simple motion models facilitate geometric analysis and yield analytical results, they often fail to capture real-world scenarios, where players' dynamics are more complex. Some studies focus on scenarios with Dubins cars and differential drive robots \cite{DC1,DC2,DC3,DDR}. In \cite{GARCIA2017139}, the active defense game with first-order missile models was addressed, providing the optimal strategy for the target. A more realistic model is the damped double integrator model. Li et al. \cite{Li} derived optimal strategies for a pursuit-evasion game with this model using the HJI framework. Reference \cite{Coon} investigated a reach-avoid game with double integrator dynamics, established capture conditions for a single defender-attacker scenario, and extended the results to multiplayer settings using a matching-based method. Wei et al. \cite{Wei} analyzed the dominance regions in an active defense game with double integrator dynamics and designed strategies based on these dominance regions. Despite these advances, reach-avoid games with damped double integrator dynamics have not been considered in existing literature.

This paper investigates optimal strategies for reach-avoid games with damped double integrator dynamics. The main contributions of this paper are as follows: (1) The reaching time of a single player is analyzed, and a multiple reachable region (MRR) is proposed. Each point within the MRR can be reached by the player using multiple distinct strategies. (2) A new type of attacker's dominance region is introduced based on the multiple reachable region. (3) Optimal strategies to reach different types of the attacker's dominance region are investigated, and it is proved that the strategies satisfy a necessary condition for optimality.

The rest of this paper is organized as follows. The reach-avoid game is formulated in Section \ref{sec:problem_formulation}. In Section \ref{sec:strategies}, we construct multiple reachable region and attacker's dominance region. Based on that, the strategies are determined and discussed. Simulation results are given in Section \ref{sec:simulations}. Finally, Section \ref{sec:conclusion} concludes the paper.

\section{Problem Formulation}\label{sec:problem_formulation}
\begin{figure}
    \centering
    \begin{tikzpicture}[scale = 1.8, font=\footnotesize]
    % 绘制坐标轴，添加方向标和x，y字母
    \draw[->] (-1.5, 0) -- (1.5, 0) node[right] {$x$}; 
    \draw[->] (0, -0.2) -- (0, 1.8) node[above] {$y$};

    % 红色三角点 (1,1) 并添加箭头标记 vA
    \draw[red!70!white, fill=red!70!white] (-1, 1.03) -- (-0.96, 0.95) -- (-1.04, 0.95) -- cycle;
    \draw[-stealth, red!70!white] (-1, 1) -- (-0.8, 0.5); % 红色箭头
    \node[left] at (-0.8, 0.5) {$\mathbf{v}_{A0}$};
    \node[left] at (-1, 1) {$\mathbf{x}_{A0}$};
    \draw[-stealth][red!70!white] (-1, 1) -- (-0.7,1.2);
    \node[above] at (-0.7,1.2) {$(u_A, \theta_A)$};
    
    \draw[red!70!white][dashed] (-1,1) .. controls (-0.9,0.75) and (-0.6,0.3) .. (-0.2,0.1);
    \draw[blue!70!white][dashed] (1,1.3) .. controls (0.6,1.6) and (0.4,1.3).. (-0.2,0.1);
    \node[right] at (0.1,0.7) {$\mathbf{x}_D(t)$};
    \node[right] at (-0.7,0.5) {$\mathbf{x}_A(t)$};
    \node at (-0.2,0.1)[star,star points=5,star point ratio=2.25,draw=red,fill=red,minimum size=2pt,scale=0.3] {};
    \node[left] at (-0.25,0.1) {$\mathbf{x}_f$};

    % 蓝色圆点 (3,2) 并添加箭头标记 vD
    \fill[blue!70!white] (1, 1.3) circle (1.5pt);
    \draw[-stealth, blue!70!white] (1, 1.3) -- (0.4, 1.7); % 蓝色箭头
    \node[right] at (0.4,1.7) {$\mathbf{v}_{D0}$};
    \node[below right] at (1,1.3) {$\mathbf{x}_{D0}$};
    \draw[-stealth][blue!70!white] (1,1.3) -- (0.9,0.8);
    \node[right] at (0.8,0.7) {$(u_D, \theta_D)$};
    
    % 绿色圆点 (0,0) 并标注 target
    \fill[green!50!black] (0, 0) circle (1.5pt);
    \node[below right] at (0, 0) {target};

\end{tikzpicture}
    \caption{An illustration of the reach-avoid game.}
    \label{fig:illstration_game}
\end{figure}
The problem considered in this paper is a reach-avoid game of one attacker (A) and one defender (D) both with damped double integrator dynamics. The attacker tries to reach the target at $[0,0]^T$ without being captured, while the defender aims to capture the attacker at the point as far as possible from the target. Fig.~\ref{fig:illstration_game} is an illustration of the game.
The states of the attacker and defender are given by their Cartesian coordinates $\mathbf{x}_A=[x_A,y_A]^T$, $\mathbf{x}_D=[x_D,y_D]^T$ and velocities $\mathbf{v}_A=[v_{Ax},v_{Ay}]^T$, $\mathbf{v}_D=[v_{Dx},v_{Dy}]^T$. The dynamics are described as follows:
\begin{equation}
\label{eq:dynamics}
    \begin{aligned}
        &\dot{x}_i = v_{ix},\\
        &\dot{y}_i = v_{iy},\\
        &\dot{v}_{ix} = -\mu v_{ix}+u_i\cos\theta_i,\\
        &\dot{v}_{iy} = -\mu v_{iy}+u_i\sin\theta_i,
    \end{aligned}
\end{equation}
where $u_i\in[0,u_{im}]$ and $\theta_i\in[0,2\pi)$ are the control inputs of player $i$, $i\in \{A,D\}$, representing the magnitude and direction of acceleration, respectively. We assume that $u_{Am}<u_{Dm}$. $\mu>0$ denotes the damping factor, which is identical for the attacker and defender. 

The terminal time is denoted as $t_f$, the terminal position of player $i$ is $\mathbf{x}_{if}=[x_{if},y_{if}]^T=[x_i(t_f),y_i(t_f)]^T$ and the terminal velocity is $\mathbf{v}_{if}=[v_{ixf},v_{iyf}]^T=[v_{ix}(t_f),v_{iy}(t_f)]^T$. Similarly, the initial position is $\mathbf{x}_{i0}$, and the velocity is $\mathbf{v}_{i0}$. 
In the defender-winning scenarios, the game terminates when the attacker is captured by the defender before reaching the target. The terminal equality constraint is
\begin{equation}\label{eq:terminal-contraint}
    g(\mathbf{x}_{Af},\mathbf{x}_{Df})=\sqrt{(x_{Af}-x_{Df})^2+(y_{Af}-y_{Df})^2}=0.
\end{equation}
And the terminal payoff function of the game is defined as
\begin{equation}\label{eq:terminal-payoff}
\begin{aligned}
    J(u_A,\theta_A,u_D,\theta_D;\mathbf{x}_{A0},\mathbf{v}_{A0},\mathbf{x}_{D0},\mathbf{v}_{D0}) = \Phi(\mathbf{x}_{Af}),
\end{aligned}
\end{equation}
where $\Phi(\mathbf{x}_{Af}) := \sqrt{x_{Af}^2+y_{Af}^2}$.

The defender and attacker are assumed to be aware of the current states of both players. The strategies of each player are mappings from the current states to the player's feasible control inputs. Denote the strategies as $u_{i}(\mathbf{x}_A,\mathbf{x}_D,\mathbf{v}_A,\mathbf{v}_D),\ \theta_i(\mathbf{x}_A,\mathbf{x}_D,\mathbf{v}_A,\mathbf{v}_D)$. The optimal strategies $u_{i}^*,\ \theta_i^*$ are defined in the sense of Nash equilibrium,
\begin{equation}
\begin{aligned}
    J(u_A^*,\theta_A^*,u_D,\theta_D)\leq J^*\leq J(u_A,\theta_A,u_D^*,\theta_D^*),\\
    \text{where }J^*=J(u_A^*,\theta_A^*,u_D^*,\theta_D^*),
\end{aligned}
\end{equation}
for all possible strategies $u_i,\ \theta_i$. The value function of the game is given as the saddle point of the terminal payoff, $\displaystyle V(\mathbf{x}_{A0},\mathbf{v}_{A0},\mathbf{x}_{D0},\mathbf{v}_{D0})=\min_{u_A,\theta_A}\max_{u_D,\theta_D}J.$

\section{Optimal Strategy in Defender-Winning Scenarios}\label{sec:strategies}

In this section, features of a single player's motion are studied and the attacker's dominance region is constructed. The optimal strategies in the defender-winning scenarios are proposed and discussed.

\subsection{Hamiltonian and Normal Solution}
\label{sec:optimal_strategy}
The Hamiltonian of this differential game is given by
\begin{equation}
\label{eq:Hamiltonian1}
\begin{aligned}
    H=&\lambda_1v_{Dx}+\lambda_2v_{Dy}+\gamma_1v_{Ax}+\gamma_2v_{Ay}\\
    +&\lambda_3(-\mu v_{Dx}+u_D\cos\theta_D)+\lambda_4(-\mu v_{Dy}+u_D\sin\theta_D)\\
    +&\gamma_3(-\mu v_{Ax}+u_A\cos\theta_A)+\gamma_4(-\mu v_{Ay}+u_A\sin\theta_A)\\
\end{aligned}
\end{equation}
where $\lambda=[\lambda_1,\lambda_2,\lambda_3,\lambda_4]^T$ is the co-state of the defender, and $\gamma=[\gamma_1,\gamma_2,\gamma_3,\gamma_4]^T$ is the co-state of the attacker.
The path equations for the defender's co-state are given by
\begin{equation}
\begin{aligned}
        &[\dot\lambda_1,\dot\lambda_2]^T=-\frac{\partial H}{\partial \mathbf{x}_D^T},\quad
        [\dot\lambda_3,\dot\lambda_4]^T=-\frac{\partial H}{\partial \mathbf{v}_D^T},\\
        &[\lambda_{1f},\lambda_{2f}]^T=\frac{\partial\Phi}{\partial\mathbf{x}_{Df}^T}+\nu\frac{\partial g}{\partial\mathbf{x}_{Df}^T},\\
        &[\lambda_{3f},\lambda_{4f}]^T=\frac{\partial\Phi}{\partial\mathbf{v}_{Df}^T}+\nu\frac{\partial g}{\partial\mathbf{v}_{Df}^T},\\
\end{aligned}
\end{equation}
where $\nu$ is a Lagrange multiplier. It can be obtained that
\begin{equation}
    \begin{aligned}
        &\dot\lambda_1=\dot\lambda_2=0,\ 
        \dot\lambda_3 = -\lambda_1+\mu\lambda_3,\ 
        \dot\lambda_4 = -\lambda_2+\mu\lambda_4,\\
        &\lambda_{3f} = \lambda_{4f} = 0,\\
        &\lambda_{1f} = \nu\frac{x_{Df}-x_{Af}}{\sqrt{(x_{Af}-x_{Df})^2+(y_{Af}-y_{Df})^2}},\\
        &\lambda_{2f} = \nu\frac{y_{Df}-y_{Af}}{\sqrt{(x_{Af}-x_{Df})^2+(y_{Af}-y_{Df})^2}}.\\
    \end{aligned}
\end{equation}
The equations of $\gamma$ are derived similarly. By solving co-states' path equations, one can obtain that
\begin{equation}
\label{eq:costateD}
\begin{aligned}
    \lambda_1 &= \nu\frac{x_{Df}-x_{Af}}{\sqrt{(x_{Af}-x_{Df})^2+(y_{Af}-y_{Df})^2}},\\
    \lambda_2 &= \nu\frac{y_{Df}-y_{Af}}{\sqrt{(x_{Af}-x_{Df})^2+(y_{Af}-y_{Df})^2}},\\
    \lambda_3 &= \frac{\lambda_1}{\mu}(1-e^{-\mu(t_f-t)}),\quad
    \lambda_4 = \frac{\lambda_2}{\mu}(1-e^{-\mu(t_f-t)}),
    \end{aligned}
\end{equation}
\begin{equation}
\label{eq:costateA}
\begin{aligned}
    \gamma_1 &= \frac{x_{Af}}{\sqrt{x_{Af}^2+y_{Af}^2}}+\nu\frac{x_{Af}-x_{Df}}{\sqrt{(x_{Af}-x_{Df})^2+(y_{Af}-y_{Df})^2}},\\
    \gamma_2 &= \frac{y_{Af}}{\sqrt{x_{Af}^2+y_{Af}^2}}+\nu\frac{y_{Af}-y_{Df}}{\sqrt{(x_{Af}-x_{Df})^2+(y_{Af}-y_{Df})^2}},\\
    \gamma_3 &= \frac{\gamma_1}{\mu}(1-e^{-\mu(t_f-t)}),\quad
    \gamma_4 = \frac{\gamma_2}{\mu}(1-e^{-\mu(t_f-t)}).
\end{aligned}
\end{equation}

The value function satisfies the Hamilton-Jacobi-Isaacs equation \cite{Isaacs}
\begin{equation}\label{eq:HJI-equation}
    \min_{u_A,\theta_A}\max_{u_D,\theta_D}\sum_{i=A,D}\left[\frac{\partial V}{\partial\mathbf{x}_i^T}\dot{\mathbf{x}_{i}}+\frac{\partial V}{\partial\mathbf{v}_i^T}\dot{\mathbf{v}}_i\right]=0,
\end{equation}
or equivalently, $H=0$, with the co-state equal to the derivative of $V$ with respect to the state. By solving $\min_{u_A,\theta_A}\max_{u_D,\theta_D}H$, we can obtain the optimal strategy
\begin{equation}
\label{eq:optimal_condition}
    \begin{aligned}
        &u_A^* = u_{Am},\ \text{if}\ \gamma_3\cos\theta_A+\gamma_4\sin\theta_A\neq 0,\\
        &\cos\theta_A^*=-\frac{\gamma_3}{\sqrt{\gamma_3^2+\gamma_4^2}},\ \sin\theta_A^*=-\frac{\gamma_4}{\sqrt{\gamma_3^2+\gamma_4^2}},\\
        &u_D^* = u_{Dm},\ \text{if}\ \lambda_3\cos\theta_D+\lambda_4\sin\theta_D\neq 0,\\
        &\cos\theta_D^*=\frac{\lambda_3}{\sqrt{\lambda_3^2+\lambda_4^2}},\ \sin\theta_D^*=\frac{\lambda_4}{\sqrt{\lambda_3^2+\lambda_4^2}}.\\
    \end{aligned}
\end{equation}

By combining \eqref{eq:costateD}, \eqref{eq:costateA} and \eqref{eq:optimal_condition}, we know that both $\theta_A^*$ and $\theta_D^*$ are constant. 
\begin{definition}[Normal strategy]
    The strategy with $u_i=u_{im}$ and a constant $\theta_i$, $i\in\{A,D\}$ is defined as Normal strategy.
\end{definition}

Then, based on \eqref{eq:dynamics} and \eqref{eq:optimal_condition}, the trajectories of the attacker and defender are given by
\begin{equation}\label{eq:velocity}
    \begin{aligned}
    v_{ix}(t) &= v_{ix0}e^{-\mu t}+\frac{u_{im}\cos\theta_i}{\mu}(1-e^{-\mu t}),\\
    v_{iy}(t) &= v_{iy0}e^{-\mu t}+\frac{u_{im}\sin\theta_i}{\mu}(1-e^{-\mu t}),    
    \end{aligned}
\end{equation}
{\vspace{-1.1cm}
\footnotesize
\begin{equation}\label{eq:motion}
    \begin{aligned}
    x_i(t) &= x_{i0}+\frac{v_{ix0}}{\mu}(1-e^{-\mu t})+\frac{u_{im}\cos\theta_i}{\mu}\left(t-\frac{1-e^{-\mu t}}{\mu}\right),\\
    y_i(t) &= y_{i0}+\frac{v_{iy0}}{\mu}(1-e^{-\mu t})+\frac{u_{im}\sin\theta_i}{\mu}\left(t-\frac{1-e^{-\mu t}}{\mu}\right).\\
    \end{aligned}
\end{equation}}
% 趋于直线，表达式

As point capture is considered, the terminal positions for the attacker and defender should be the same, denoted as $\mathbf{x}_{f}=\mathbf{x}_{Af}=\mathbf{x}_{Df}$. The terminal time and optimal $\theta_i^*$ can be determined through \eqref{eq:motion} once the terminal position is given. 

In the defender winning scenarios, the defender captures the attacker before it reaches the target. So the attacker chooses the optimal terminal position in the region where it can reach before the defender, named the attacker's dominance region. The aim of the following content is to determine the terminal position $\mathbf{x}_f$ by constructing the attacker's dominance region.

\subsection{Multiple Reachable Region}\label{sec:single}
The isochrone at $t$ of player $i,\ i\in\{A,D\}$, denoted as $\mathcal{I}_i(t)$, is the set of points where player $i$ can reach at $t$ with Normal strategy. 
\begin{lemma}\label{lemma:reachable-region}
    Given $\mathbf{x}_{i0}$ and $\mathbf{v}_{i0}$, the isochrone is a circle with radius $r_{ic}(t)$ and center $\mathbf{x}_{ic}(t)=[x_{ic}(t),y_{ic}(t)]^T$ given by
    \begin{equation}\label{eq:isochrones}
        \begin{aligned}
            r_{ic}(t) &= \frac{u_{im}}{\mu}\left(t-\frac{1-e^{-\mu t}}{\mu}\right),\\
            x_{ic}(t) &= x_{i0}+\frac{v_{ix0}}{\mu}(1-e^{-\mu t}),\\
            y_{ic}(t) &= y_{i0}+\frac{v_{iy0}}{\mu}(1-e^{-\mu t}).
        \end{aligned}
    \end{equation}
    The set $\{\mathbf{x}\in\mathbb{R}^2|\|\mathbf{x}-\mathbf{x}_{ic}(t)\|_2\leq r_{ic}(t)\}$ represents all points that player $i$ can reach at $t$ with feasible control input.
\end{lemma}

The proof is given in Appendix \ref{sec:ap-reach}.

Initially, two players' isochrones are separated. After they are externally tangent, they subsequently intersect. Since $u_{Dm}>u_{Am}$, the defender's isochrone eventually encloses the attacker's after internal tangency. At this stage, capture is guaranteed regardless of the attacker's strategy. Throughout the subsequent analysis, we assume that all considered scenarios occur before the first internal tangency. Now, we show that the isochrones are tangent only finitely many times.
\begin{theorem}\label{thm:circumscribe}
Isochrones $\mathcal{I}_A(t)$ and $\mathcal{I}_D(t)$ are externally tangent / internally tangent for at least once and at most three times.
\end{theorem}

The proof is given in Appendix \ref{sec:ap-tangent}. 

\begin{corollary}\label{cor:reach_point}
    Any point $\mathbf{x}\in\mathbb{R}^2$ can be reached by a player via at least one and at most three Normal strategies.
\end{corollary}
\begin{proof}
    Consider the case where $u_{Am}=0$, $\mathbf{v}_{A0}=[0,0]^T$, and $\mathbf{x}_{A0}=\mathbf{x}$. In this case, $\mathcal{I}_A(t)$ degenerates to the point $\mathbf{x}$; thus, the condition that $\mathcal{I}_D(t)$ passes through $\mathbf{x}$ is equivalent to saying that $\mathcal{I}_D(t)$ is tangent to $\mathcal{I}_A(t)$. 
    By Theorem~\ref{thm:circumscribe}, $\mathcal{I}_D(t)$ intersects $\mathbf{x}$ at least once and at most three times. It follows that the defender has at least one and at most three Normal strategies to reach $\mathbf{x}$. The case for the attacker can be treated similarly.
\end{proof}

\begin{definition}[Multiple Reachable Region, MRR]
    The multiple reachable region of player $i$, designated by $\mathcal{M}_i$, $i\in\{A,D\}$, is the set of points in $\mathbb{R}^2$ that the player can reach through more than one Normal strategy.
\end{definition}
An illustration of $\mathcal{M}_i$ is shown in Fig.~\ref{fig:region}. There are three distinct times for the player to reach a certain position $\mathbf{x}$ in $\mathcal{M}_i$. 
\begin{figure}
    \centering
    \subfigure[]{
        \input{fig/region}
        \label{fig:region}}
    \subfigure[]{
        \input{fig/reaching_time_exp}
        \label{fig:reaching-time-iso}}
    \caption{An illustration of isochrones and multiple reachable region $\mathcal{M}_i$. The isochrones are marked by colored circles, and the region $\mathcal{M}_i$ is shaded in Fig~\ref{fig:region}. }
    \label{fig:region-and-isochrones}
\end{figure}

\begin{definition}[Reaching Time]
    Given initial states $\mathbf{x}_{A0}$ and $\mathbf{v}_{A0}$, the attacker's reaching time at $\mathbf{x}$ is the time instant when the attacker with Normal strategy reaches $\mathbf{x}$, denoted as $t_{Aj}(\mathbf{x};\mathbf{x}_{A0},\mathbf{v}_{A0})$, where $j\in\{1,2,3\}$ for $\mathbf{x}\in\mathcal{M}_A$ and $j=0$ for $\mathbf{x}\in\mathbb{R}^2$\textbackslash$\mathcal{M}_A$. Reaching times for the defender are defined similarly as $t_{Dk}(\mathbf{x},\mathbf{x}_{D0},\mathbf{v}_{D0})$.
    The order of magnitude of player $i$'s reaching times in $\mathcal{M}_i$ is $t_{i1}(\mathbf{x})<t_{i2}(\mathbf{x})<t_{i3}(\mathbf{x})$.
\end{definition}
The reaching time outside $\mathcal{M}_i$ is abbreviated as $t_i$ hereinafter. Clearly, the isochrone $\mathcal{I}_i(t)$ passes through $\mathbf{x}$ at reaching time $t_{ij}(\mathbf{x})$, as presented in Fig.~\ref{fig:reaching-time-iso}. In comparison, the isochrones of a simple motion player are concentric circles; therefore, the MRR does not exist for a simple motion player. This is a distinct difference between the two models, which leads to different properties of the attacker's dominance region.

\begin{lemma}\label{lemma:unreachable}
    For player $i$, $i\in\{A,D\}$, a point $\mathbf{x}\in\mathcal{M}_i$ is unreachable under any strategy during the interval $(t_{i2}(\mathbf{x}),t_{i3}(\mathbf{x}))$, but is reachable during the interval $(t_{i1}(\mathbf{x}),t_{i2}(\mathbf{x}))$.
\end{lemma}
\begin{proof}
    From the proof of Theorem~\ref{thm:circumscribe} and Corollary~\ref{cor:reach_point}, we know that for $t\in(t_{i2}(\mathbf{x}),t_{i3}(\mathbf{x}))$, $\|\mathbf{x}-\mathbf{x}_{ic}(t)\|_2 > r_{ic}(t)$, which implies that $\mathbf{x}$ lies outside the region enclosed by the isochrone. By Lemma~\ref{lemma:reachable-region}, $\mathbf{x}$ cannot be reached at such $t$. Similarly, for $t\in(t_{i1}(\mathbf{x}),t_{i2}(\mathbf{x}))$, $\mathbf{x}$ lies inside the region enclosed by the isochrone, so the player has strategies to reach $\mathbf{x}$ during this period. 
\end{proof}
\begin{lemma}\label{lemma:t-change}
    As $u_{im}$, $i\in\{A,D\}$, decreases, $t_{i2}$ decreases, while $t_{ij}$, $j\in\{0,1,3\}$, increases.
\end{lemma}
\begin{proof}
    As $u_i$ decreases, the center of the isochrone is unchanged, while the radius given by \eqref{eq:isochrones} decreases. From Lemma~\ref{lemma:unreachable}, the isochrone passes through $\mathbf{x}\in\mathcal{M}_i$ at $t_{ij}$. Due to the decreased radius, the point enters the region enclosed by the isochrone later and exits earlier; consequently, $t_{i1}$ and $t_{i3}$ increase, while $t_{i2}$ decreases. By the same reasoning, the case $\mathbf{x}\in\mathbb{R}^2\setminus\mathcal{M}_i$ can be established.
\end{proof}
\begin{lemma}\label{lemma:t-crit}
    Reaching times $t_{ij}(\mathbf{x})$, $j\in\{0,1,3\}$ satisfy $v_{ix}\cos\theta_i+v_{iy}\sin\theta_i>0$, while the reaching time $t_{i2}(\mathbf{x})$ satisfies $v_{ix}\cos\theta_i+v_{iy}\sin\theta_i<0$,     where $v_{ix},\ v_{iy}$ are the velocities of player $i$ when it reaches $\mathbf{x}$ at $t_{ij}(\mathbf{x})$, and $\theta_i$ is the corresponding strategy. 
\end{lemma}
\begin{proof}
    After $t_{ij}(\mathbf{x})$, $j\in\{0,1,3\}$, the position $\mathbf{x}$ is inside the region enclosed by the isochrone $\mathcal{I}_i$, while after $t_{i2}(\mathbf{x})$, $\mathbf{x}$ is outside the region enclosed by the isochrone $\mathcal{I}_i$. Write out the time derivative of the radius of isochrone minus the distance between $\mathbf{x}$ and $\mathbf{x}_{ic}$
    \begin{equation}\label{eq:reaching-time-crit}
    \begin{aligned}
        \frac{d}{dt}(r_{ic}-\|\mathbf{x}-\mathbf{x}_{ic}\|_2)=\frac{u_i}{\mu}(1-e^{-\mu t})+\frac{e^{-\mu t}\mathbf{v}^T_{i0}(\mathbf{x}-\mathbf{x}_{ic})}{r_{ic}}\\
        =\frac{u_i}{\mu}(1-e^{-\mu t})+e^{-\mu t}\mathbf{v}^T_{i0}[\cos\theta_i,\sin\theta_i]^T\\
        =v_{ix}\cos\theta_i+v_{iy}\sin\theta_i.\\
    \end{aligned}
    \end{equation}
    Therefore, \eqref{eq:reaching-time-crit} is greater than $0$ for $t_{ij}(\mathbf{x})$, $j\in\{0,1,3\}$, and smaller than $0$ for $t_{i2}$. 
\end{proof}
\begin{lemma}
    The gradient of reaching time $t_{ij}(\mathbf{x})$, $i\in\{A,D\}$, $j\in\{0,1,2,3\}$ is
    \begin{equation}\label{eq:dtdx}
    \frac{dt_{ij}(\mathbf{x})}{d\mathbf{x}}=\frac{1}{v_{ix}\cos\theta_i+v_{iy}\sin\theta_i}[\cos\theta_i,\sin\theta_i]^T,
    \end{equation}
    where $v_{ix},\ v_{iy}$ are the velocities of player $i$ when it reaches $\mathbf{x}$ at $t_{ij}(\mathbf{x})$, and $\theta_i$ is the corresponding strategy. 
\end{lemma}
\begin{proof}
     The player's trajectory $\mathbf{x}_i$ is a function of $\theta_i$ and $t$. The gradient of $t$ can be obtained from the derivative of $\mathbf{x}_i$ using the inverse function theorem. The Jacobian matrix of $\mathbf{x}_i(\theta_i,t)$ in \eqref{eq:motion} is given by
    \begin{equation}\label{eq:Jacobi}
        \frac{\partial (x_i,y_i)}{\partial(\theta_i,t)}=\begin{bmatrix}
            \frac{\partial x_i}{\partial\theta_i}&\frac{\partial x_i}{\partial t}\\
            \frac{\partial y_i}{\partial\theta_i}&\frac{\partial y_i}{\partial t}\\
        \end{bmatrix}
        =
        \begin{bmatrix}
            -r_{ic}\sin\theta_i&v_{ix}\\
            r_{ic}\cos\theta_i&v_{iy}\\
        \end{bmatrix}.
    \end{equation}
    The gradient of reaching time is obtained by
    \begin{equation}
    \begin{aligned}
        &\frac{\partial(\theta_i,t)}{\partial (x_i,y_i)}=\left(\frac{\partial (x_i,y_i)}{\partial(\theta_i,t)}\right)^{-1}=\\
        &-\frac{1}{r_{ic}(v_{ix}\cos\theta_i+v_{iy}\sin\theta_i)}
        \begin{bmatrix}
            v_{iy}&-v_{ix}\\
            -r_{ic}\cos\theta_i&-r_{ic}\sin\theta_i\\
        \end{bmatrix}.
    \end{aligned}
    \end{equation}
\end{proof}

\begin{definition}[Self Intersection Point]
    Given $\mathcal{I}_i(t)$ intersecting with $\mathcal{I}_i(t+\Delta t)$, self intersection points $\mathbf{x}^\pm_{int}(\Delta t,t)$ are the intersections of $\mathcal{I}_i(t)$ and $\mathcal{I}_i(t+\Delta t)$, where $(\mathbf{x}^+_{int}(\Delta t,t)-\mathbf{x}_{i0})\wedge\mathbf{v}_{i0}\geq0$ and $(\mathbf{x}^-_{int}(\Delta t,t)-\mathbf{x}_{i0})\wedge\mathbf{v}_{i0}\leq0$ ($\wedge$ is the wedge product of two vectors).
\end{definition}
All points in $\mathcal{M}_i$ are self intersection points, and $t$ and $\Delta t$ are determined by reaching times, for example, $\mathbf{x}\in\mathcal{M}_i$ is self intersection point $\mathbf{x}^\pm_{int}(t_{i2}(\mathbf{x})-t_{i1}(\mathbf{x}),t_{i1}(\mathbf{x}))$.
When $\Delta t = 0$, the two isochrones coincide, so the self intersection points do not exist. However, for sufficiently small but non-zero $\Delta t$, they exist. These points are instrumental in analyzing the reaching times and the MRR.
\begin{lemma}\label{lemma:self-overlap}
    At self intersection points $\lim_{\Delta t\to 0}\mathbf{x}^\pm_{int}(\Delta t,t)$, $v_{ix}\cos\theta_i+v_{iy}\sin\theta_i=0$, where $\mathbf{v}_i$ is the velocity when reaching at the point and $\theta_i$ is the strategy.
\end{lemma}
\proof
    For simplicity, assume $v_{iy0}=0$,\ $\mathbf{x}_{i0}=[0,0]^T$. For the case $v_{iy0}\neq 0$, we could rotate the coordinate to have $v_{iy0}=0$. For brevity, $\lim_{\Delta t\to 0}\mathbf{x}^\pm_{int}(\Delta t,t)$ is denoted as $\mathbf{x}^\pm_{int}(t)$ in the following part. The $x$-coordinate of $\mathbf{x}^\pm_{int}(t)$ is given by the solution $x$ of the following equations
    \begin{equation}\label{eq:solve-self-intersect}
        \begin{aligned}
            &(x-x_{ic}(t))^2+y^2=r_{ic}^2(t),\\
            &(x-x_{ic}(t+\Delta t))^2+y^2=r_{ic}^2(t+\Delta t).
        \end{aligned}
    \end{equation}
    Isochrone's radius and $x$-coordinate \eqref{eq:isochrones} are
    \begin{equation}\label{eq:Isochron-Taylor}
        \begin{aligned}
            r_{ic}(t+\Delta t) &= r_{ic}(t)+\frac{u_{im}}{\mu}(1-e^{-\mu t})\Delta t+o(\Delta t),\\
            x_{ic}(t+\Delta t) &= x_{ic}(t)+v_{ix0}e^{-\mu t}\Delta t+o(\Delta t).
        \end{aligned}
    \end{equation}
    Taking \eqref{eq:Isochron-Taylor} into \eqref{eq:solve-self-intersect} gives
    \begin{equation}
        (x-x_{ic}(t))v_{ix0}e^{-\mu t}=-r_{ic}(t)\frac{u_{im}}{\mu}(1-e^{-\mu t})+\frac{o(\Delta t)}{\Delta t}.
    \end{equation}
    By taking the limit as $t$ approaches 0, we obtain $\mathbf{x}^\pm_{int}(t)=x_{ic}(t)-r_{ic}(t)\frac{u_i}{\mu}\frac{1-e^{-\mu t}}{v_{ix0}e^{-\mu t}}$.
    The corresponding strategy is given by 
    \begin{equation}\label{eq:overlap-control}
        \cos\theta_i=\frac{x^\pm_{int}(t)-x_{ic}(t)}{r_{ic}(t)}=-\frac{u_{im}}{\mu}\frac{1-e^{-\mu t}}{v_{ix0}e^{-\mu t}}.
    \end{equation}
    By substituting \eqref{eq:overlap-control} and \eqref{eq:velocity} into \eqref{eq:dtdx}, we see that $v_{ix}\cos\theta_i+v_{iy}\sin\theta_i=0$. 
\endproof
The intersections $\lim_{\Delta t\to 0}\mathbf{x}^\pm_{int}(\Delta t,t)$ exist.
Because $\mathcal{I}_i(t)$ and $\lim_{\Delta t\to 0}\mathcal{I}_i(t+\Delta t)$ pass through the intersections simultaneously, the corresponding reaching times are equal. So, the set $\{\lim_{\Delta t\to 0}\mathbf{x}^\pm_{int}(\Delta t,t)\}$ is the set of points where two of the three reaching times equal. 
What's more, $v_{ix}\cos\theta_i+v_{iy}\sin\theta_i$ equals zero, thus the gradient \eqref{eq:dtdx} is not defined at the intersections. The set $\{\lim_{\Delta t\to 0}\mathbf{x}^\pm_{int}(\Delta t,t)\}$ forms a curve. On the one side of the curve, there are three different reaching times, and on the other side of the curve, two equal reaching times do not exist.
Thus, the curve is the boundary of multiple reachable region. 
Moreover, $t_i(\mathbf{x}),\ i\in\{A,D\}$ is continuous in $\mathbb{R}^2$\textbackslash$\mathcal{M}_i$. 
$t_{i1}(\mathbf{x}),t_{i2}(\mathbf{x}),t_{i3}(\mathbf{x})$ are continuous in $\mathcal{M}_i$. 
Depending on which two reaching times are equal, the boundary is categorized into two types: $\mathcal{B}^i_{I}=\{\mathbf{x}\in\mathbb{R}^2|t_{i1}(\mathbf{x})=t_{i2}(\mathbf{x})\}$ and $\mathcal{B}^i_{II}=\{\mathbf{x}\in\mathbb{R}^2|t_{i2}(\mathbf{x})=t_{i3}(\mathbf{x})\}$, as shown in Fig.~\ref{fig:region}. On $\mathbf{B}_{I}^i$, the reaching time surface $t_{i3}(\mathbf{x})$ coincides with $t_i(\mathbf{x})$, and similarly, $t_{i1}=t_i$ on $\mathbf{B}_{II}^i$. Therefore, the reaching times $t_{i3}$ and $t_{i1}$ are continuous across the boundary $\mathcal{B}_I^i$ and $\mathcal{B}_{II}^i$, respectively, while $t_{i2}$ is discontinuous at boundary points because its gradient is not defined.

The boundary of MRR can be constructed based on the Lemma.~\ref{lemma:self-overlap}. By combining
\begin{equation}
    \begin{aligned}
        &v_{ix}\cos\theta_i+v_{iy}\sin\theta_i=0,\\
        &\cos^2\theta_i^2+\sin^2\theta_i=1,
    \end{aligned}
\end{equation}
we obtain the Normal strategy to reach the boundary
\begin{equation}\label{eq:boundary}
\begin{aligned}
    \sin\theta_i=\frac{-v_{iy0}\frac{u_i}{\mu}(e^{\mu t}-1)\pm v_{ix0}\sqrt{v_{i0}^2-\frac{u_i^2}{\mu^2}(e^{\mu t}-1)^2}}{v_{i0}^2},\\
    \cos\theta_i=\frac{-v_{ix0}\frac{u_i}{\mu}(e^{\mu t}-1)\mp v_{iy0}\sqrt{v_{i0}^2-\frac{u_i^2}{\mu^2}(e^{\mu t}-1)^2}}{v_{i0}^2}.
\end{aligned}
\end{equation}
Substituting \eqref{eq:boundary} into \eqref{eq:motion} yields the position of the boundary formed at $t$.

Now, we introduce two more concepts that help the construction and analysis of MRR and reaching time. Firstly, the barrier time refers to the moment after which player’s isochrones do not self-overlap anymore\cite{Li}, which can be expressed as
\begin{equation}\label{eq:ts}
    t_{s}=\frac{1}{\mu}\ln{\frac{\mu\sqrt{v_{ix0}^2+v_{iy0}^2}+u_{im}}{u_{im}}},\ i\in\{A,D\}.
\end{equation}
We have omitted the subscript $i$ in $t_s$.

Secondly, it can be observed from Fig.~\ref{fig:region} that the tangent vector of the boundary does not exist at two points $\mathbf{x}_{u1}$ and $\mathbf{x}_{u2}$. 
We investigate the time when the player's isochrones pass $\mathbf{x}_{u1}$, which is essential for constructing $\mathcal{M}_i$ using self-overlapping intersections. Denote $\mathbf{x}(t)$ as the point on $\partial\mathcal{M}_i$ with the parameter $t$ referring to the two equal reaching times. The tangent vector of $\partial\mathcal{M}_i$ at $\mathbf{x}(t)$ can be derived based on \eqref{eq:motion}:
\begin{equation}
    \frac{d \mathbf{x}}{d t}=\mathbf{v}_{i}(t,\theta_i)+\frac{u_i}{\mu}(t-\frac{1-e^{-\mu t}}{\mu})\frac{d \theta_i}{d t}[-\sin\theta_i,\cos\theta_i]^T,
\end{equation}
where $\theta_i$ is given by \eqref{eq:boundary}.
We get $\frac{d\theta_i}{d t}$ as
\begin{equation}
\begin{aligned}
    \frac{d\theta_i}{d t}=\frac{1}{\cos\theta_i}\frac{d \sin\theta_i}{dt}
    =\pm\frac{e^{\mu t}u_i}{M},
\end{aligned}
\end{equation}
where $M=\sqrt{v_{ix0}^2+v_{iy0}^2-\frac{u_i^2}{\mu^2}(e^{\mu t}-1)^2}$.
On the boundary, $\mathbf{v}_i(t,\theta_i)$ is parallel to $[-\sin\theta_i,\cos\theta_i]^T$. Therefore, the tangent vector is initially in the same direction with $\mathbf{v}_i(t,\theta_i)$, then equals to $\mathbf{0}$ at $\mathbf{x}_{u1}$ and $\mathbf{x}_{u2}$. After that, it is in the opposite direction with the velocity. The time $t_{u}:=t_{i2}(\mathbf{x}_{u1})=t_{i2}(\mathbf{x}_{u2})$ is determined by solving $\frac{d\mathbf{x}}{dt}=\mathbf{0}$:
\begin{equation}
    \begin{aligned}
        \frac{u_i}{\mu}\left(t_u-\frac{1-e^{-\mu t_u}}{\mu}\right)\frac{e^{\mu t_u}u_i}{M}&=\|\mathbf{v_i}(t_u,\theta_i)\|_2,\\
        &=Me^{-\mu t_u}.
        \end{aligned}
\end{equation}

Period $[0,t_{s}]$ can be divided into two phases: $[0,t_{u}]$ and $[t_{u},t_{s}]$. 
The initial position $\mathbf{x}_{i0}\in\mathcal{B}^i_I$, because $t_{i1}(\mathbf{x}_{i0})=t_{i2}(\mathbf{x}_{i0})=0$, and $t_{i3}(\mathbf{x}_{i0})>0$ if $\|\mathbf{v}_{i0}\|_2>0$. 
As for $\mathbf{x}_{s}$, since $\lim_{\Delta t\to 0}\mathbf{x}^\pm_{int}(\Delta t,t_{s})=\mathbf{x}_{s}$ and $\mathcal{I}_i(t)$ do not pass $\mathbf{x}_{s}$ after $t_{s}$, $t_{i2}(\mathbf{x}_s)=t_{i3}(\mathbf{x}_s)$. Thus, $\mathbf{x}_s$ is on $\mathcal{B}^i_{II}$.
In the first phase, $t_{i1}=t_{i2}$, corresponding to $\mathcal{B}^i_I$. In the second phase, $t_{i2}=t_{i3}$, corresponding to $\mathcal{B}^i_{II}$. On $\mathcal{B}^i_I$, as $t_{i2}$ increases from $0$ to $t_{u}$, $t_{i3}$ decreases to $t_{u}$. On $\mathcal{B}^i_{II}$, as $t_{i2}$ increases from $t_{u}$ to $t_{s}$, $t_{i1}$ decreases from $t_{u}$. At $\mathbf{x}_{u1}$ and $\mathbf{x}_{u2}$, $t_{i1}=t_{i2}=t_{i3}$. We could distinguish different reaching time by Lemma~\ref{lemma:t-crit},
\begin{equation}
    \begin{aligned}
        &j=0,\quad\text{if}\ t_{ij}(\mathbf{x})>t_s,\\
        &j=1,\quad\text{if}\ t_{ij}(\mathbf{x})<t_u,\ v_{ixf}\cos\theta_i+v_{iyf}\sin\theta_i>0,\\
        &j=2,\quad\text{if}\ t_{ij}(\mathbf{x})<t_s,\ v_{ixf}\cos\theta_i+v_{iyf}\sin\theta_i<0,\\
        &j=3,\quad\text{if}\ t_u<t_{ij}(\mathbf{x})<t_s,\ v_{ixf}\cos\theta_i+v_{iyf}\sin\theta_i>0.
    \end{aligned}
\end{equation}

\begin{figure}[htbp]
    \centering
    \scalebox{0.55}{\input{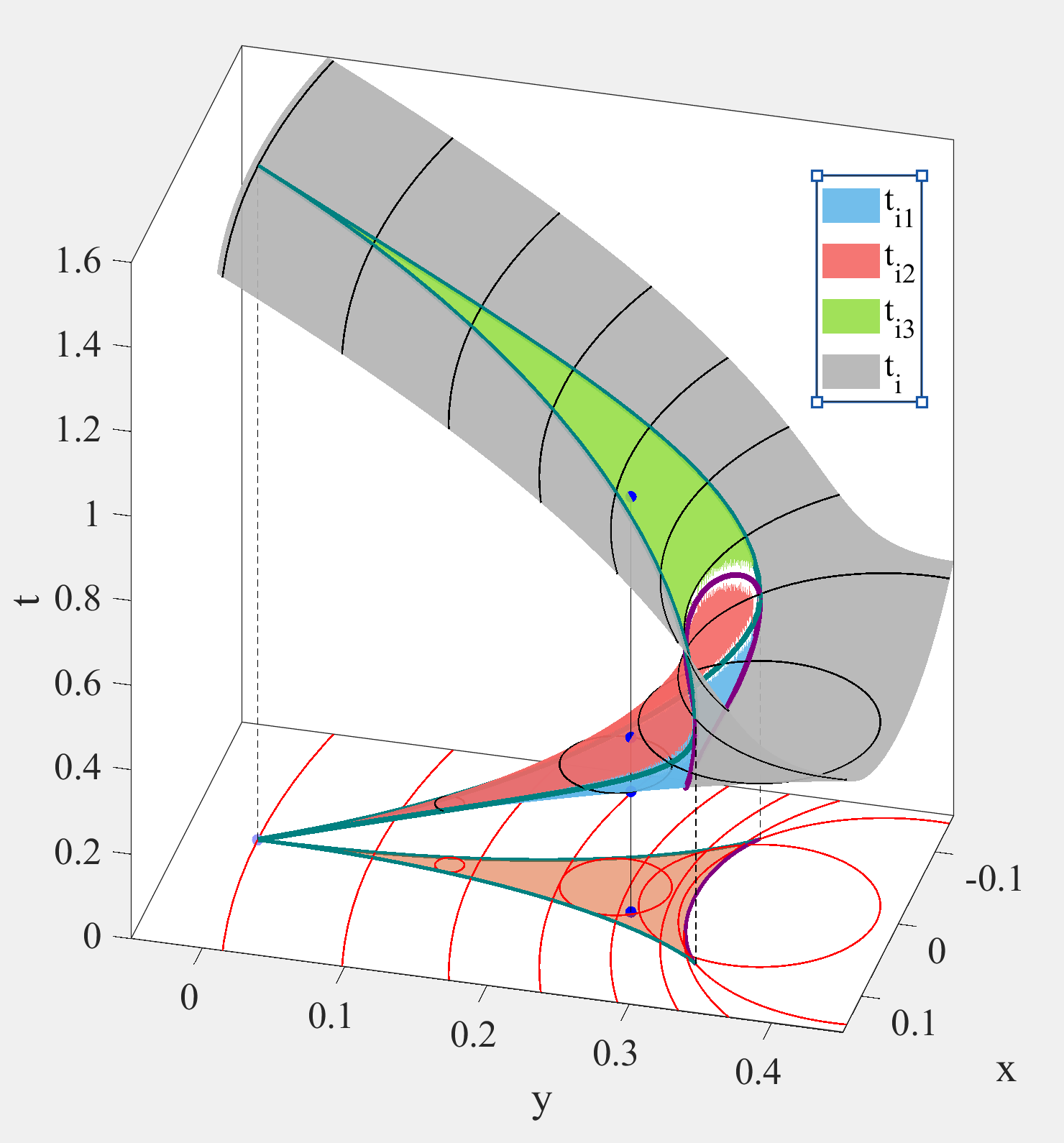}}
    \caption{An illustration of the reaching times of player $i$. The surfaces represent reaching times $t_{ij}(\mathbf{x})$. The black circles are isochrones. The projection on $x-y$ plane is given.}
    \label{fig:reaching-time}
\end{figure}

Fig.~\ref{fig:reaching-time} is an illustration of the reaching times and multiple reachable region. 

\subsection{Attacker's Dominance Region}

As defined in \cite{Isaacs}, the attacker's dominance region is the region where the attacker can reach without being captured. In this section, we first define the normal attacker's dominance region for this problem. Then, a new type of the attacker's dominance region is introduced.
\begin{definition}[ADR]\label{def:ADR}
    The normal attacker's dominance region is the set of points where the attacker can reach with Normal strategy when it is outside the region enclosed by $\mathcal{I}_D$.
    \begin{equation}
    \begin{aligned}
        \text{ADR}=\{&\mathbf{x}\in\mathbb{R}^2\mid \exists j\in\{0,1,2,3\},\\
        &\|\mathbf{x}-\mathbf{x}_{Dc}(t_{Aj}(\mathbf{x}))\|_2>r_{Dc}(t_{Aj}(\mathbf{x}))\}.
    \end{aligned}
    \end{equation}
    
    ADR is further divided into two types. The first one is the region where the attacker can reach without being captured, given by 
    \begin{equation}
        \begin{aligned}
            \mathcal{R}_I = \{&\mathbf{x}\in\mathbb{R}^2\mid
            \exists j\in\{0,1,2,3\},\ \forall t<t_{Aj},\\
            &\|\mathbf{x}_A(\theta_A,u_{Am},t)-\mathbf{x}_{Dc}(t)\|_2>r_{Dc}(t)\},
        \end{aligned}
    \end{equation}
    where $\theta_A$ is the strategy for the attacker to reach $\mathbf{x}$ at $t_{Aj}$.
    The second type of ADR is $\mathcal{R}_{II}=$ ADR\textbackslash$\mathcal{R}_I$, which means that there exists a defender's strategy to capture the attacker. 

    % The curve formed by the intersections of isochrones is denoted as $\mathcal{L}$, The boundary of ADR consists of $\mathcal{L}$ and $\mathbf{x}\in\mathcal{B}^A_I$.  
\end{definition}
\begin{remark}
For damped double integrator dynamics, the optimal trajectory is not always a straight line, so the attacker arriving earlier than the defender at point $\mathbf{x}$ does not guarantee the attacker arriving earlier along the attacker's whole trajectory. Therefore, the second type of the attacker's dominance region exists. In comparison, a simple motion player moves along a straight line, so $\mathcal{R}_{II}$ does not exist.
\end{remark}

We denote by $\mathcal{L}$ the intersections of the attacker's and defender's isochrones prior to the first internal tangency. On $\mathcal{L}$, the reaching times of the attacker and defender equal. 
Moreover, on two sides of $\mathcal{B}_I^A$, the relationship of the reaching times could be $t_{A1}<t_{D}<t_{A3}$ and $t_{A}>t_{D}$ respectively. This means the attacker may reach $\mathcal{M}_A$ before the defender and reach $\mathbb{R}^2$\textbackslash$\mathcal{M}_A$ after the defender. However, the attacker will not be captured on $\mathcal{B}_{I}^A$ because the attacker's and defender's reaching times are not equal along $\mathcal{B}_{I}^A$. Therefore, the points on $\mathcal{B}_{I}^A$ are not feasible terminal positions and will not be discussed here. $\mathcal{L}$ forms the boundary of $\mathcal{R}_I\cup\mathcal{R}_{II}$.

According to Lemma~\ref{lemma:unreachable}, for the attacker to safely reach the terminal position, $\mathcal{R}_{I}\cup\mathcal{R}_{II}$ is the set of points where there exists some $j$ such that $t_{Aj}(\mathbf{x})<\min_k t_{Dk}(\mathbf{x})$ or $t_{D2}(\mathbf{x})<t_{Aj}(\mathbf{x})<t_{D3}(\mathbf{x})$. The region where $t_{D1}(\mathbf{x})<t_{Aj}(\mathbf{x})<t_{D2}(\mathbf{x})$ is not included in $\mathcal{R}_{I}\cup\mathcal{R}_{II}$. However, based on Lemma~\ref{lemma:t-change}, the attacker can change its reaching time by adopting a smaller $u_A$, so it is possible for the attacker to reach a point $\mathbf{x}\in\mathcal{M}_D$ during $(t_{D2},t_{D3})$ if it can reach the point during $(t_{D1},t_{D2})$ with Normal strategy. 
Therefore, another type of the attacker's dominance region should exist inside $\mathcal{M}_D$.

\begin{definition}\label{def:thirdADR}
The third type of the attacker's dominance region $\mathcal{R}_{III}$ consists of points $\mathbf{x}\in\mathcal{M}_D$ for which there exists $j\in\{0,1,3\}$ such that $t_{D1}(\mathbf{x})<t_{Aj}(\mathbf{x})<t_{D2}(\mathbf{x})$, and there exist $u_A(t)$, $\theta_A(t)$ and $t_f$ satisfying that $\mathbf{x}_A(t_f)=\mathbf{x}$, where $\mathbf{x}$ lies outside the region enclosed by $\mathcal{I}_D(t_f)$, and $t_f$ is prior to the time of the first internal tangency.
\end{definition}

According to Lemma~\ref{lemma:t-change}, when the attacker adopts a smaller $u_A$, the reaching time $t_{Aj}(\mathbf{x}),\ j\in\{0,1,3\}$ increases and exceeds $t_{D2}$. As a result, there exists $\hat{u}_A<u_{Am}$ such that the attacker reaches the point $\mathbf{x}$ in $\mathcal{R}_{III}$ at the time instant $t_{D2}(\mathbf{x})$.
However, there are regions in $\mathcal{M}_D$ satisfying $t_{D1}(\mathbf{x})<t_{Aj}(\mathbf{x})<t_{D2}(\mathbf{x})$ but cannot be reached by the attacker safely. As the attacker decreases $u_A$, the first internal tangency time decreases. And the region cannot be reached if $t_{D2}(\mathbf{x})$ is bigger than the decreased internal tangency time. To categorize the regions in $\mathcal{M}_D$ and find $\mathcal{R}_{III}$, we first investigate the properties of $\mathcal{L}$, which are useful for us to judge the relation between $t_{Aj}$ and $t_{Dk}$.

Let $\mathcal{L}$ intersects with the boundary of $\mathcal{M}_D$ at $\mathbf{x}$, and $t$ is the time $\mathcal{I}_A$ and $\mathcal{I}_D$ intersect at $\mathbf{x}$. Boundary $\mathcal{L}$ has the following properties:
\begin{enumerate}
    \item If $\mathbf{x}\in\mathcal{B}_I^D$ and $t=t_{D3}(\mathbf{x})$ or $\mathbf{x}\in\mathcal{B}_{II}^D$ and $t=t_{D1}(\mathbf{x})$, $\mathcal{L}$ penetrates the boundary because $t_{D3}$ is continuous at any point on $\mathcal{B}_{I}^D$ and $t_{D1}$ is continuous at any point on $\mathcal{B}_{II}^D$.
    \item If $t=t_{D2}(\mathbf{x})$, $\mathcal{L}$ is tangent to $\mathcal{M}_D$'s boundary. This is because $t_{i2}$ is not continuous at boundary points.
\end{enumerate}
Due to the continuity of $t_{Aj}(\mathbf{x})$ and $t_{Dk}(\mathbf{x})$, the subscripts $j,k$ of $t_{Aj}(\mathbf{x}) = t_{Dk}(\mathbf{x})$ change on $\mathbf{x}\in\mathcal{L}$ only if $\mathcal{L}$ intersects with the boundary of $\mathcal{M}_A$ or $\mathcal{M}_D$. 

\begin{lemma}\label{lemma:dta=dtd}
Assume that $t_{Aj}=t_{Dk}$ on part of $\mathcal{L}$, then $t_{Aj}<t_{Dk}$ on one side and $t_{Aj}>t_{Dk}$ on the other side, i.e., the order of magnitude changes.
\end{lemma}
\begin{proof}
Let the tangent vector of $\mathcal{L}$ at $\mathbf{x}$ be $\mathbf{b}$, while the normal vector $\mathbf{n}$ is perpendicular to $\mathbf{b}$. Parameterize $\mathcal{L}$ by $s$.
The reaching times of the attacker and defender are equal along $\mathcal{L}(s)$, so
\begin{equation}\label{eq:equal-grad}
    \frac{dt_{Dk}(\mathbf{x})}{d s}=\frac{dt_{Aj}(\mathbf{x})}{d s},
\end{equation}
where $\mathbf{x}\in\mathcal{L}$. The derivative can be expressed as $\frac{dt_{ij}(\mathbf{x})}{d s}=\frac{d\mathbf{x}}{d s}\frac{dt_{ij}(\mathbf{x})}{d\mathbf{x}}=\mathbf{b}^T\frac{dt_{ij}(\mathbf{x})}{d\mathbf{x}}$.
Therefore, \eqref{eq:equal-grad} can be rewritten as
\begin{equation}
    \mathbf{b}^T\frac{dt_{Dk}(\mathbf{x})}{d\mathbf{x}}=\mathbf{b}^T\frac{dt_{Aj}(\mathbf{x})}{d\mathbf{x}}.    
\end{equation}

If the relative magnitude order between $t_{Aj}$ and $t_{Dk}$ remains unchanged, we may assume that $t_{Aj}(\mathbf{x})=t_{Dk}(\mathbf{x})$ and $t_{Aj}(\mathbf{x}+\beta\mathbf{n})>t_{Dk}(\mathbf{x}+\beta\mathbf{n})$ on both sides of $\mathcal{L}$, for $|\beta|$ smaller than a given constant. 
Then, because $\mathbf{x}$ is the local minimal point of $t_{Aj}(\mathbf{x})-t_{Dk}(\mathbf{x})$ along direction $\mathbf{n}$, we can derive that
\begin{equation}
    \mathbf{n}^T\frac{dt_{Dk}(\mathbf{x})}{d\mathbf{x}}=\mathbf{n}^T\frac{dt_{Aj}(\mathbf{x})}{d\mathbf{x}}.    
\end{equation}

Therefore, $\frac{dt_{Dk}(\mathbf{x})}{d\mathbf{x}}=\frac{dt_{Aj}(\mathbf{x})}{d\mathbf{x}}$ at any points on $\mathcal{L}$. However, it can be verified from \eqref{eq:dtdx} that $\frac{dt_{Aj}(\mathbf{x})}{d\mathbf{x}}$ and $\frac{dt_{Dk}(\mathbf{x})}{d\mathbf{x}}$ are parallel if and only if $\mathcal{I}_A$ and $\mathcal{I}_D$ are tangent, which is satisfied for finite times, meaning that $\frac{dt_{Dk}(\mathbf{x})}{d\mathbf{x}}=\frac{dt_{Aj}(\mathbf{x})}{d\mathbf{x}}$ for finite times. Hence, on two sides separated by $\mathcal{L}$, the order of the attacker's and defender's reaching times differs.
\end{proof}

Now, we introduce two types of region in $\mathcal{M}_D$ that belong to the third type of the attacker's dominance region $\mathcal{R}_{III}$.
\begin{figure}
    \centering
    \subfigure[]{
        \scalebox{0.8}{\input{fig/ADR3z}}
    \label{subfig:egADR3}}
    \subfigure[]{
        \scalebox{0.8}{\input{fig/ADR3s}}
    \label{subfig:egADR1}}
    \subfigure[]{
        \scalebox{0.8}{\input{fig/counter-example-RIII}}
    \label{subfig:egADR2}}
    \caption{Illustrations for $\mathcal{R}_{III}$. The yellow region represents the ADR. The pink and orange regions represent MRR, where the orange region corresponds to the overlap between the MRR and the ADR.}
\end{figure}
\begin{theorem}\label{thm:thirdADR}
    The region that satisfies either of the following two conditions is the third type of the attacker's dominance region. 
    \begin{enumerate}
        \item Region $\mathcal{P}_1$: the boundary of $\mathcal{P}_1$ consists solely of intersections of isochrones between the first and second isochrones external tangent times $t_{o1}$ and $t_{o2}$, and $\|\mathbf{x}_{Ac}(t)-\mathbf{x}_{Dc}(t)\|_2>r_{Dc}(t)$ during $[t_{o1},t_{o2}]$.
        \item Region $\mathcal{P}_2$: the boundary of $\mathcal{P}_2$ consists solely of $\mathcal{B}_{II}^D$ and part of $\mathcal{L}$ where $t_{Aj}(\mathbf{x})=t_{D2}(\mathbf{x})$, $j\in\{0,1,3\}$.
    \end{enumerate}
\end{theorem}
\begin{proof}
    For simplicity, we use $t_A(\mathbf{x})$ to refer to attacker's reaching times $t_{Aj}(\mathbf{x})$, $j\in\{0,1,3\}$.
    We first discuss region $\mathcal{P}_1$, as shown in Fig.~\ref{subfig:egADR3}. 
    As $u_A$ decreases, the center of $\mathcal{I}_A(t)$ remains unchanged, while the radius decreases. Consequently, the first external tangency occurs later but the second external tangency occurs earlier. Moreover, during the time period $[t_{o1}(u_{Am}),t_{o2}(u_{Am})]$, since $\|\mathbf{x}_{Ac}(t)-\mathbf{x}_{Dc}(t)\|_2>r_{Dc}(t)$, $\mathcal{I}_A$ and $\mathcal{I}_D$ do not intersect if $u_A=0$, indicating that no internal tangency occurs. There exists a value $\hat{u}_A$ such that $\mathcal{I}_A$ is externally tangent to $\mathcal{I}_D$ only once at the time $t_{o1}(\hat{u}_A)=t_{o2}(\hat{u}_A)$. For $u_A<\hat{u}_A$, $\mathcal{I}_A$ and $\mathcal{I}_D$ do not intersect over $[t_{o1}(u_{Am}),t_{o2}(u_{Am})]$; therefore, the region $\mathcal{P}_1$ contracts to a point as $u_A$ decreases from $u_{Am}$ to $\hat{u}_A$. Therefore, for any position $\mathbf{x}\in\mathcal{P}_1$, there exists $u_A\in[\hat{u}_A,u_{Am}]$ such that the attacker reaches at $\mathbf{x}$ when it is outside the region enclosed by $\mathcal{I}_D$. 

    If $t_A(\mathbf{x})>t_{D3}(\mathbf{x})$ for $\mathbf{x}\in\mathcal{P}_1$, $\mathbf{x}$ is contained in $\mathcal{I}_D(t)$ for $t>t_A(\mathbf{x})$, which is contradictory to the existence of $\hat{u}_A$ given in the last paragraph. Therefore, $t_{D1}<t_{A}<t_{D2}$ inside $\mathcal{P}_1$. $\mathcal{P}_1$ is the third type of the attacker's dominance region. 
    
    Next, we analyse the region $\mathcal{P}_2$, as shown in Fig.~\ref{subfig:egADR1}. The intersections of $\mathcal{B}_{II}^D$ and $\mathcal{L}$ are tangent points $\mathbf{x}_{tg1}$ and $\mathbf{x}_{tg2}$, where $t_{A}(\mathbf{x}_{tgi})=t_{D2}(\mathbf{x}_{tgi})=t_{D3}(\mathbf{x}_{tgi})$. If $t_A\equiv t_{D2}$ on $\mathcal{L}$, $\mathcal{L}$ would only be tangent to the boundary of $\mathcal{M}_D$ and could not penetrate the region. Hence, there must be pairs of tangent points $\mathbf{x}_{tgi}$. Between tangent points, $t_A=t_{D2}$, before $\mathbf{x}_{tg1}$ and after $\mathbf{x}_{tg2}$, $t_A=t_{D3}$ so that it can exit $\mathcal{M}_D$. Since the neighboring attacker's dominance region satisfies $t_{D2}(\mathbf{x})<t_A(\mathbf{x})<t_{D3}(\mathbf{x})$, from Lemma~\ref{lemma:dta=dtd}, $t_{D1}(\mathbf{x})<t_A(\mathbf{x})<t_{D2}(\mathbf{x})$ in $\mathcal{P}_2$. 

    $\mathbf{x}_{tg1}$ and $\mathbf{x}_{tg2}$ vary continuously as $u_A$ decreases. Otherwise, $\mathbf{x}_{tg1}$ would either vanish or exhibit a jump discontinuity before the condition $\mathbf{x}_{tg1}(u_A)=\mathbf{x}_{tg2}(u_A)$ is reached. In that case, all points in $\mathcal{P}_2$ would abruptly change from $t_A(\mathbf{x})<t_{D2}(\mathbf{x})$ to $t_A(\mathbf{x})>t_{D2}(\mathbf{x})$, implying that $t_A(\mathbf{x})$ in $\mathcal{P}_2$ is not continuous with respect to $u_A$. This contradicts the continuity of $t_A(\mathbf{x})$. 
    According to Lemma~\ref{lemma:t-change}, $t_{A}(\mathbf{x})$ increase as $u_A$ decreases. So, $\mathbf{x}_{tgi}$ moves on $\mathcal{B}_{II}^D$ into parts that $t_A<t_{D2}$, which is between $\mathbf{x}_{tg1}$ and $\mathbf{x}_{tg2}$. Similarly, the part of $\mathcal{L}$ where $t_A=t_{D2}$ contracts into $\mathcal{P}_2$ as $u_A$ decreases.   
    Consequently, $\mathcal{P}_2$ continuously contracts to a point as $u_A$ decreases. Hence, for any position $\mathbf{x}\in\mathcal{P}_2$, there exists $u_A<u_{Am}$ such that the attacker reaches at $\mathbf{x}$ when it is outside the region enclosed by $\mathcal{I}_D$. 
    Because $t_{A}$ along $\mathcal{L}$ decreases from $t_{D3}$ to $t_{D2}$ at $\mathbf{x}_{tg1}$ and increases at $\mathbf{x}_{tg2}$, the minimum intersection time of isochrones---corresponding to the external tangency---occurs on $\mathcal{L}$ between $\mathbf{x}_{tg1}$ and $\mathbf{x}_{tg2}$. As $u_A$ decreases, a new internal tangency may occur at $t$ when the radius of $\mathcal{I}_A(t)$ shrinks such that the two intersection points between $\mathcal{I}_A(t)$ and $\mathcal{I}_D(t)$ coalesce into a single point. But isochrones intersect at the boundary of $\mathcal{P}_2$ prior to the internal tangency time. This is because the external tangency between $\mathbf{x}_{tg1}$ and $\mathbf{x}_{tg2}$ occurs earlier, and the time on $\mathcal{L}$ varies continuously. Moreover, since $\mathcal{P}_2$ contracts to a point, all the points in $\mathcal{P}_2$ can be reached before internal tangency. To aid understanding, a counterexample is given in Fig.~\ref{subfig:egADR2}. The region $t_{D1}(\mathbf{x})<t_A(\mathbf{x})<t_{D2}(\mathbf{x})$ is denoted by $\mathcal{P}$. Its boundary has two parts, $t_{A}(\mathbf{x})=t_{D2}(\mathbf{x})$ and $t_{A}(\mathbf{x})=t_{D1}(\mathbf{x})$ (denoted as $\pmb{l}$), which does not satisfy the conditions in Theorem~\ref{thm:thirdADR}. The boundaries $\mathcal{L}$ corresponding to different values of $u_A$ are illustrated: the black, grey and white curves represent $u_A=1$, $u_A=0.4$, and $u_A=0.3$, respectively. During the decrease of $u_A$ from $0.4$ to $0.3$, a new internal tangency occurs on $\pmb{l}$. As a result, part of the boudary formed after the internal tangency is excluded, leading to a smaller region enclosed by $\mathcal{L}$ at $u_A=0.3$ compared with that at $u_A=0.4$.
  
\end{proof}

\begin{figure}
    \centering
    \input{fig/changeR3}
    \caption{The yellow region represents the ADR. The pink region $\mathcal{R}_{III}$ contracts in the direction indicated by arrows as $u_A$ decreases. $\mathcal{L}$ under different $u_A$ are color curves. Two separate $\mathcal{R}_{III}$ regions have formed at $u_A=0.9$. $\mathcal{P}_1$ has vanished at $u_A=0.7$, and $\mathcal{P}_2$ keeps contracting with the tangent points $\mathbf{x}_{tgi}$ moving towards each other on $\mathcal{B}_{II}^D$ until they merge. }
    \label{fig:changeR3}
\end{figure}
An example of how $\mathcal{R}_{III}$ changes with $u_A$ is presented in Fig.~\ref{fig:changeR3}. 

\begin{remark}
    Compared to \cite{Coon} and \cite{Wei}, the influence of reaching times are considered for the first time in this paper. $\mathcal{R}_{III}$ is introduced to extend the normal attacker's dominance region into a larger area, so that the attacker can reach the outside of the normal attacker's dominance region with a smaller $u_A$.
\end{remark}

\begin{figure}
    \centering
    \subfigure[Attacker's dominance region]{
        \input{fig/ADR}
    \label{subfig:ADR}}
    \subfigure[Zoom of Fig.~\ref{subfig:ADR}]{
        \input{fig/ADR2}
    \label{subfig:ADR2}}\\
    \caption{An example of the attacker's dominance region. The yellow region represents the ADR, i.e., $\mathcal{R}_I\cup\mathcal{R}_{II}$. The pink region represents $\mathcal{R}_{III}$.}
    \label{fig:ADR}
\end{figure}

We provide illustrations of the attacker's dominance region in Fig.~\ref{fig:ADR}. In this example, isochrones are externally tangent three times before being internally tangent. The area isolated in $\mathcal{R}_{I}\cup\mathcal{R}_{II}$ satisfies the first condition given in Theorem~\ref{thm:thirdADR}, while the other $\mathcal{R}_{III}$ region satisfies the second condition.

\subsection{Optimal Strategy}\label{sec:discuss_strategy}
The terminal position $\mathbf{x}_f$ is the closest point to the target in the attacker's dominance region.
We propose two strategies regarding different terminal position.
\begin{enumerate}
    \item \textbf{ADR strategy}: $\mathbf{x}_f\in\mathcal{L}$. Reach $\mathbf{x}_f$ with $u_i=u_{im}$ and constant $\theta_i$.\\
    \item \textbf{MRR strategy}: $\mathbf{x}_f\in\mathcal{R}_{III}$. Reach $\mathbf{x}_f$ with constant $\theta_i$ and $u_D=u_{Dm}$, $u_A=\hat{u}_A$ determined by $\frac{\hat{u}_A}{\mu}\left(t_{D2}(\mathbf{x})-\frac{1-e^{-\mu t_{D2}(\mathbf{x})}}{\mu}\right)=\|\mathbf{x}-\mathbf{x}_{Ac}(t_{D2}(\mathbf{x}))\|_2$ in order to reach the point at the same time with the defender.
\end{enumerate}
The strategy is optimal if HJI equation \eqref{eq:HJI-equation} is fulfilled for all initial states \cite{Isaacs}. Next, we demonstrate that $H=0$ for MRR strategy and in some situations for ADR strategy, which provides a necessary condition to be optimal.

\begin{theorem}\label{thm:second-optimal-condition}
    $H(\mathbf{x}_i,\theta_i^*,u_i^*,\lambda,\gamma,t)=0$ holds along the trajectory for ADR strategy if $t_{Dk}(\mathbf{x}_f)=t_{Aj}(\mathbf{x}_f)$, $k\in\{0,1,3\},\ j\in\{0,1,3\}$ or $t_{D2}(\mathbf{x}_f)=t_{A2}(\mathbf{x}_f)$, and holds for MRR strategy.
\end{theorem}
\begin{proof}
    We first prove that $H=0$ holds for ADR strategy with given constraints. The optimal position $\mathbf{x}_f$ on $\mathcal{L}$ minimizes the distance to the target: 
    \begin{equation}\label{eq:minimal-condition-re}
    \begin{aligned}
        \frac{d\|\mathbf{x}_f\|_2}{d\mathbf{x}_f}
        +\sigma\left(\frac{\partial t_A}{\partial \mathbf{x}_f}-\frac{\partial t_D}{\partial \mathbf{x}_f}\right)=0,
    \end{aligned}
    \end{equation}
    where $\sigma$ is the Lagrange multiplier, $t_A$ and $t_D$ are reaching times to $\mathbf{x}_f$.
    Define guessed position co-states $\hat\lambda$ and $\hat\gamma$ by
    \begin{equation}\label{eq:guess_costate}
        \begin{aligned}
            \hat\lambda=[\hat\lambda_1,\hat\lambda_2]^T&=-\sigma\frac{\partial t_D(\mathbf{x}_f,\mathbf{x}_{D0},\mathbf{v}_{D0})}{\partial \mathbf{x}_{D0}},\\
            \hat\gamma=[\hat\gamma_1,\hat\gamma_2]^T&=\sigma\frac{\partial t_A(\mathbf{x}_f,\mathbf{x}_{A0},\mathbf{v}_{A0})}{\partial \mathbf{x}_{A0}}.
        \end{aligned}
    \end{equation}
    By \eqref{eq:motion}, $\frac{\partial t_i(\mathbf{x}_f,\mathbf{x}_{i0},\mathbf{v}_{i0})}{\partial \mathbf{x}_{i0}}=-\frac{\partial t_i(\mathbf{x}_f,\mathbf{x}_{i0},\mathbf{v}_{i0})}{\partial \mathbf{x}_f}$. The guessed velocity co-state is $[\hat{\lambda}_3,\hat{\lambda}_4]=\frac{1-e^{-\mu(t_f-t)}}{\mu}\hat\lambda$, $[\hat{\gamma}_3,\ \hat{\gamma}_4]=\frac{1-e^{-\mu(t_f-t)}}{\mu}\hat\gamma$.
    
    We now verify that this reach-avoid game's co-state in \eqref{eq:Hamiltonian1} is the guessed co-state, when $t_A$ and $t_D$ are reaching times $t_{ij}$ both with $j\in\{0,1,3\}$ or both with $j=2$. To this end, we demonstrate that guessed co-state satisfies co-state's terminal constraints and path equations. Then, we use $\hat\lambda$ and $\hat\gamma$ to analyze the condition $H=0$ \eqref{eq:Hamiltonian1}.
    
    Position co-state's terminal constraints are $[\lambda_1,\lambda_2]=\frac{\partial\Phi}{\partial\mathbf{x}_D^T}+\nu\frac{\partial g}{\partial\mathbf{x}_D^T}$, $[\gamma_1,\gamma_2]=\frac{\partial\Phi}{\partial\mathbf{x}_A^T}+\nu\frac{\partial g}{\partial\mathbf{x}_A^T}$, which gives
    \begin{equation}\label{eq:terminal-relation}
        \lambda_1+\gamma_1=\frac{x_f}{\sqrt{x_f^2+y_f^2}},\ \lambda_2+\gamma_2=\frac{y_f}{\sqrt{x_f^2+y_f^2}}.
    \end{equation}
    It can be inferred from \eqref{eq:minimal-condition-re} that $\hat\lambda$ and $\hat\gamma$ also satisfy this relationship.
    The velocity co-state's terminal constraint is $\lambda_3=\lambda_4=\gamma_3=\gamma_4=0$, which is satisfied by the guessed co-state. 

    Solving $\min_{\theta_A}\max_{\theta_D}H(\mathbf{x}_{i},\theta_i,u_i,\hat{\lambda},\hat{\gamma},t)$ gives the strategy $\hat{\theta}_i$. 
    Reformulate this as $\displaystyle\hat{\theta}_A=\arg\min_{\theta_A}[\cos\theta_A,\sin\theta_A]\hat{\gamma}$, $\displaystyle\hat{\theta}_D=\arg\max_{\theta_D}[\cos\theta_D,\sin\theta_D]\hat{\lambda}$, equivalent to
    \begin{equation}\label{eq:path-eqn-condition}
        \hat{\theta}_i=\arg\min_{\theta_i}\sigma\frac{\partial t_i(\mathbf{x}_f,\mathbf{x}_{i0},\mathbf{v}_{i0})}{\partial \mathbf{x}_{i0}^T}[\cos\theta_i,\sin\theta_i]^T.
    \end{equation}
        This solves an optimization problem with cost function $J=\text{sgn}(\sigma)t_f=\pm t_f$, denoted as the auxiliary problem. $\text{sgn}(\cdot)$ is the sign function.
    To see this, we write out the Hamiltonian of the auxiliary problem
    \begin{equation*}
    \begin{aligned}
      H_{aux}=&\kappa_{i1}v_{ix}+\kappa_{i2}v_{iy}\\+&\kappa_{i3}(-\mu v_{ix}+u_i\cos\theta_i)+\kappa_{i4}(-\mu v_{iy}+u_i\sin\theta_i).
    \end{aligned}
    \end{equation*}
    The co-state path equations are $[\dot{\kappa}_{i1},\dot{\kappa}_{i2}]=-\frac{\partial H_{aux}}{\partial \mathbf{x}_i^T}=0,\ [\dot{\kappa}_{i3},\dot{\kappa}_{i4}]=-\frac{\partial H_{aux}}{\partial \mathbf{v}_i^T}=-[\kappa_{i1},\kappa_{i2}]^T+\mu[\kappa_{i3},\kappa_{i4}]^T$. So, $[\kappa_{i3},\kappa_{i4}]=\frac{(1-e^{-\mu (t_f-t)})}{\mu}[\kappa_{i1},\kappa_{i2}]$.
    According to \cite{Isaacs}, $[\kappa_{i1},\kappa_{i2}]=\pm\frac{\partial t_i(\mathbf{x},\mathbf{x}_{i}(t),\mathbf{v}_{i}(t))}{\partial \mathbf{x}_{i}^T(t)}$. Therefore, the auxiliary problem's optimal strategy $\theta_{aux}=\arg\min_{\theta}H_{aux}$ is equivalent to \eqref{eq:path-eqn-condition}.
    
    Solving the auxiliary problem gives local minimal (maximal) reaching time because it minimizes $J=\pm t_f$. If both players reach $\mathbf{x}_f$ at a local minimal (maximal) reaching time, the optimal strategy $\theta_i$ satisfies \eqref{eq:path-eqn-condition}. In this case, the guessed co-states determine the optimal strategy. Because the guessed velocity co-state shares the same form as the co-state in \eqref{eq:costateA} and \eqref{eq:costateD}, and the guessed position co-state is constant, they satisfy the path equations.
    By contrast, if one player reaches $\mathbf{x}_f$ at a local minimal time while the other one reaches it at a local maximal, the optimal strategies $\theta_A$ and $\theta_D$ cannot simultaneously satisfy \eqref{eq:path-eqn-condition}.
    Solution $t_{i2}(\mathbf{x})$ is a local maximal time to reach $\mathbf{x}$ because for $t\in[t_{i1}(\mathbf{x}),t_{i2}(\mathbf{x})]$, $\mathbf{x}$ is in the region enclosed by isochrones, but for $t\in[t_{i2}(\mathbf{x}),t_{i3}(\mathbf{x})]$, it is outside the region enclosed the isochrones. Solutions $t_{ij},\ j\in\{0,1,3\}$ are local minimal reaching time. 
    
    At $t=t_f$, $\lambda_3=\lambda_4=0$ and $\gamma_3=\gamma_4=0$, so $H=\lambda_1v_{Dxf}+\lambda_2v_{Dyf}+\gamma_1v_{Axf}+\gamma_2v_{Ayf}$. 
    By substituting \eqref{eq:dtdx} into \eqref{eq:guess_costate} we can obtain that
    \begin{equation}
        H=\hat\gamma^T\mathbf{v}_{Af}+\hat\lambda^T\mathbf{v}_{Df}=0.
    \end{equation}
    Because $\frac{dH}{dt}=0$, for local minimal point $\mathbf{x}_f\in\mathcal{L}$ where $t_{Dk}(\mathbf{x})=t_{Aj}(\mathbf{x})$, $k\in\{0,1,3\},\ j\in\{0,1,3\}$ or $t_{D2}(\mathbf{x})=t_{A2}(\mathbf{x})$, $H=0$ is satisfied along the trajectory.

    An analogous proof establishes for the MRR strategy, if we suppose the attacker is a slower player with $u_{Am}=\hat{u}_A$. Now we explain why $u_A=\hat{u}_A$ is consistent with the Maximum Principle, which means $u_{Am}=\arg\min_{u_A}H$ for optimal strategy. This can be explained by the condition that the Hamiltonian \eqref{eq:Hamiltonian1} is independent of $u_A$, i.e., $\gamma_1\cos\theta_A+\gamma_2\sin\theta_A=0$. For terminal position $\mathbf{x}_f\in\mathcal{R}_{III}$, it is in the MRR. Though changing $\mathbf{x}_{A0}$ will influence the shape of $\mathcal{L}$, the MRR is unchanged, and the terminal position is still optimal. Because $\hat{u}_A<u_{Am}$, the attacker can still find a new $u_A$ to reach $\mathbf{x}_f$ at $t_{D2}(\mathbf{x}_f)$. Therefore, the final cost remains invariant, the derivative of the value function satisfies $\frac{\partial V}{\partial\mathbf{x}_{A0}}=0$, equivalently, $\gamma_1=\gamma_2=0$. So the Hamiltonian is independent of $u_A$. The terminal position for MRR strategy satisfies $t_{Aj}(\mathbf{x}_f)=t_{D2}(\mathbf{x}_f)$, $j\in\{0,1,3\}$. Though the attacker and defender's reaching times are not both minimal (maximal) reaching time, the co-states $\gamma_1$ and $\gamma_2$ are zero, so $H=0$ still holds. 
    
\end{proof}

If $\mathbf{x}_f\in\mathcal{L}$ and $t_{D2}=t_{Aj},\ j\in\{0,1,3\}$, or $t_{Dk}=t_{A2},\ k\in\{0,1,3\}$, $H\neq 0$ along the trajectory. (1) For the case $t_{D2}=t_{Aj},\ j\in\{0,1,3\}$, the boundary $\mathcal{L}$ shifts outwards the attacker's dominance region if the attacker adopts smaller $u_A$, so $\|\mathbf{x}_f\|_2$ deceases, which means ADR strategy is not optimal under this situation. And such position $\mathbf{x}_f$ is on the boundary of $\mathcal{R}_{III}$ so it can be improved by adopting MRR strategy. (2) If $t_{D1}=t_{A2}$ or $t_D=t_{A2}$, then $t_{A1}<\min t_{D}$ on both sides of $\mathcal{L}$, so both sides are in $\mathcal{R}_{I}\cup\mathcal{R}_{II}$.
(3) If $t_{D3}=t_{A2}$, the side where $t_{D2}<t_{A2}<t_{D3}$ is in $\mathcal{R}_I\cup\mathcal{R}_{II}$ by Definition~\ref{def:ADR}, while the other side where $t_{D3}<t_{A2}$ may be outside the ADR. However, $t_{A2}$ decreases as the attacker adopts a smaller $u_A$, which means $\mathcal{L}$ moves into region that $t_{D3}<t_{A2}$, so the attacker has a better target point by adopting smaller $u_A$. 

Whenever the ADR strategy fails to satisfy the necessary condition, the performance can be improved by reducing $u_A$. In this case, the corresponding terminal point lies adjacent to a new type of the attacker’s dominance region, and the MRR strategy should be adopted, under which the necessary condition is satisfied. Theorem~\ref{thm:thirdADR} identifies two types of region, but there are still other cases remaining unsolved such as the one in Fig.~\ref{subfig:egADR2}. In such cases, changing $u_A$ can improve the outcome, but the shape of the attacker's dominance region and the optimal terminal position is not determined yet.

\section{Simulations}\label{sec:simulations}
This section conducts three experiments to demonstrate the effectiveness of our strategies. The parameters are set as follow: $u_A=1,u_D=2,\mu=1$, the target is at $[0,0]^T$. 

In the first experiment, the minimal point is in $\mathcal{R}_I$. The attacker and defender adopt ADR strategy, denoted as $\theta_i^*$. For comparison, we use pure-pursuit strategy, i.e. $\theta_A(t)=\arctan\frac{y_{A}(t)}{x_{A}(t)}$ and $\theta_D(t)=\arctan\frac{y_{A}(t)-y_{D}(t)}{x_{A}(t)-x_{D}(t)}$. We also use the optimal strategy for the pursuit-evasion game in \cite{Li}, denoted as optimal-pursuit. The attacker will check whether it can reach the target without being captured. Once the attacker can, it will adopt the corresponding strategy to reach the target.
The initial states are set as follow: $x_{A0}=1,\ y_{A0}=-0.1,\ x_{D0}=1.2,\ y_{D0}=0.1,\ v_{Ax0}=0,\ v_{Ay0}=0,\ v_{Dx0}=-0.5,\ v_{Dy0}=-1$. The results are presented in Fig.~\ref{fig:example1}. 
If both the attacker and the defender apply ADR strategy, the optimal terminal point $\mathbf{x}_f^*$ does not change according to time, so the trajectory overlap with the open-loop strategy's trajectory, as shown in Fig.~\ref{subfig:1trajectory}. From Table~\ref{tab:distance} we can see that ADR strategy is the optimal, and if a player changes its strategy unilaterally, the terminal cost will become worse.
\begin{figure*}
    \centering
    \vspace{-0.35cm} %设置与上面正文的距离
    \subfigtopskip=2pt %设置子图与上面正文或别的内容的距离
    \subfigbottomskip=2pt %设置第二行子图与第一行子图的距离，即下面的头与上面的脚的距离
    \subfigcapskip=-5pt %设置子图与子标题之间的距离
    \subfigure[($\theta_D^*$,$\theta_A^*$)]{
        \input{fig/example2/test}
    \label{subfig:1trajectory}}
    \subfigure[($\theta_D^*$,pure-pursuit)]{
        \input{fig/example2/test-pureA}
    \label{subfig:1trajectory-pureA}}
    \subfigure[(pure-pursuit,$\theta_A^*$)]{
        \input{fig/example2/test-pureD}
    \label{subfig:1trajectory-pureD}}\\
    \subfigure[($\theta_D^*$,optimal-pursuit)]{
        \input{fig/example2/test_chaseA}
    \label{subfig:1trajectory-chaseA}}
    \subfigure[(optimal-pursuit,$\theta_A^*$)]{
        \input{fig/example2/test_chaseD}
    \label{subfig:1trajectory-chaseD}}
    \caption{Simulations of case 1. The yellow region represents the ADR. The target is marked with green dot. Terminal positions are stars. Attacker and defender's initial positions are triangle and circle, respectively. The trajectories of the attacker and defender applying open loop ADR strategy are red and blue lines, the attacker's and defender's trajectories of different strategies are pink and green dashed lines. }
    \label{fig:example1}
\end{figure*}

\begin{table}[htbp]
    \centering
    \renewcommand{\arraystretch}{1.05}
    \setlength{\tabcolsep}{4pt}
    \scriptsize
    \caption{Final distance to target}
    \begin{tabular}{ccc}
        \toprule
        Defender's strategy& Attacker's strategy&Distance \\
        \hline
        $\theta_D^*$&$\theta_A^*$&0.79\\
        $\theta_D^*$&pure-pursuit&0.86\\
        pure-pursuit&$\theta_A^*$&0\\
        $\theta_D^*$&optimal-pursuit&1.56\\
        optimal-pursuit&$\theta_A^*$&0.72\\
        \bottomrule
    \end{tabular}
    \label{tab:distance}
\end{table}

In the second experiment, the real minimal point is in $\mathcal{R}_{III}$ and ADR strategy does not work. The initial states are set as follow: $x_{A0}=-0.3457,\ y_{A0}=0.0517,\ x_{D0}=-0.6728,\ y_{D0}=-0.0455,\ v_{Ax0}=0.0862,\ v_{Ay0}=0.0338,\ v_{Dx0}=1.6534,\ v_{Dy0}=0.0907$. We first simulate the situation that both the attacker and the defender ignore $\mathcal{R}_{III}$. That is, both the attacker and the defender adopt ADR strategy to reach the minimal point in $\mathcal{R}_{I}\cup\mathcal{R}_{II}$. As a result, the optimal point changes with time, which is illustrated in Fig.~\ref{fig:example4}. 
\begin{figure*}
    \centering
    % \vspace{-0.35cm} %设置与上面正文的距离
    \subfigtopskip=2pt %设置子图与上面正文或别的内容的距离
    \subfigcapskip=5pt %设置子图与子标题之间的距离
    \subfigure[initial states]{
        \input{fig/example4/test0}
    \label{subfig:4overview}}
    \subfigure[$t=0s$]{
        \input{fig/example4/test1}
    \label{subfig:4t=0}}
    \subfigure[$t=0.045s$]{
        \input{fig/example4/test2}
    \label{subfig:4t=0.045}}
    \subfigure[$t=0.1075s$]{
        \input{fig/example4/test3}
    \label{subfig:4t=0.1075}}
    \caption{Simulations of case 2. (a), (b) The initial states and optimal point. (c) When $t=0.045s$, the optimal point changes from previous position (grey star) to another position (green star). The boundaries of ADR from $t=0s$ to $0.045s$ are curves in gray to black colors. (d) When $t=0.1075s$, the optimal point changes again. The boundaries of ADR from $t=0.045s$ to $0.1075s$ are curves in gray to black colors. }
    \label{fig:example4}
\end{figure*}

We then simulate the situation that the attacker adopts MRR strategy to reach the optimal point in $\mathcal{R}_{III}$, while the defender reaches the same point in maximum acceleration. As shown in Fig.~\ref{subfig:4s-AD-R3}, two players reach the point at the same time. If the defender still applies ADR strategy, the payoff function will become even smaller when the game terminates, presented in Fig.~\ref{subfig:4s-A-R3}.
\begin{figure}
    \centering
    \vspace{-0.1cm} %设置与上面正文的距离
    \subfigtopskip=2pt %设置子图与上面正文或别的内容的距离
    \subfigcapskip=5pt %设置子图与子标题之间的距离
    \subfigure[]{
        \input{fig/example4/test_R3_corD}
    \label{subfig:4s-AD-R3}}\hspace{0.5cm}
    \subfigure[]{
        \input{fig/example4/test_R3}
    \label{subfig:4s-A-R3}}
    \caption{Simulations of case 2. (a) The attacker and defender both adopt MRR strategy. (b) The attacker adopts MRR strategy, while the defender adopts ADR strategy.}
    \label{fig:example4s}
\end{figure}

In the third case, the attacker passes through the $\mathcal{R}_{II}$ to reach the terminal position. The attacker applies ADR strategy while defender aims to capture the attacker in the shortest time, i.e. defender's target point is the intersection of attacker's trajectory and boundary of $\mathcal{R}_{III}$. The initial states are set as follow: $x_{A0}=0,\ y_{A0}=-5,\ x_{D0}=-0.5,\ y_{D0}=-4.9,\ v_{Ax0}=0,\ v_{Ay0}=0,\ v_{Dx0}=2,\ v_{Dy0}=0$. The result of the game are shown in Fig.~\ref{fig:example5}. The attacker neglects the possibility of being captured, resulting in the attacker being captured before arriving expected terminal point.
\begin{figure}
    \centering
    \vspace{-0.35cm} %设置与上面正文的距离
    \subfigtopskip=2pt %设置子图与上面正文或别的内容的距离
    \subfigbottomskip=2pt %设置第二行子图与第一行子图的距离，即下面的头与上面的脚的距离
    \subfigcapskip=-5pt %设置子图与子标题之间的距离
    \subfigure[trajectory]{
        \input{fig/test}
    \label{subfig:trajectory}}
    \\
    \subfigure[]{
        \input{fig/test2}
    \label{subfig:detail}}
    \caption{Simulations of case 3 that the attacker has to cross $\mathcal{R}_{III}$ to reach the optimal point in $\mathcal{R}_{II}$. The attacker adopts strategy I, while the defender tries to capture the attacker as soon as possible. As a result, the attacker is captured before reaching the optimal point. The boundaries of ADR at different times are curves with colors from grey to black. }
    \label{fig:example5}
\end{figure}

\section{Conclusion}\label{sec:conclusion}
In this article, a reach-avoid differential game with two damped double integrator players has been addressed. We analysed the multiple reachable region and introduced a new type of attacker's dominance region. We highlighted that the attacker's dominance region has distinct properties compared to other dynamic models, such as simple motion. Strategies to reach certain parts in the attacker's dominance region were proposed and were proved to satisfy the necessary condition for the strategy to be optimal. Finally, performance of the proposed strategies in various scenarios was verified by numerical simulations. 

Strategies in the following three scenarios are worth further investigation. (1) The shape of the third type of the attacker's dominance region is not completely determined yet. (2) If the attacker has to cross the second type of the attacker's dominance region to reach the terminal position, the defender can capture the attacker at some intermediate location. (3) The dominance region may not be convex, there are situations involving multiple minimal points. 

\appendices
\section{Proof of Lemma~\ref{lemma:unreachable}}\label{sec:ap-reach}
\proof
    Let $\mathbf{v}_i(t)=\mathbf{v}_{i0}e^{-\mu t}+\mathbf{q}(t)$, where $\mathbf{q}(t)=[q_x(t),q_y(t)]^T$ satisfies
    \begin{equation}\label{eq:q}
        \begin{aligned}
            &\dot{q}_x(t)=-\mu q_x(t)+u_i(t)\cos\theta_i(t),\\
            &\dot{q}_y(t)=-\mu q_y(t)+u_i(t)\sin\theta_i(t),\\
            &q_x(0)=0,\ q_y(0)=0,
        \end{aligned}
    \end{equation}
    with $u_i(t)<u_{im}$. The trajectory of the player is given by:
    \begin{equation}
        \mathbf{x}_i(t)=\mathbf{x}_{i0}+\frac{\mathbf{v}_{i0}}{\mu}(1-e^{-\mu t})+\int_0^t\mathbf{q}(\tau)d\tau.
    \end{equation}
    Then we have
    \begin{equation}\label{eq:xr}
        \|\mathbf{x}_i(t)-\mathbf{x}_{ic}(t)\|_2=\left\Vert\int_0^t\mathbf{q}(\tau)d\tau\right\Vert_2\leq \int_0^t\|\mathbf{q}(\tau)\|_2d\tau.
    \end{equation}
    
    According to \eqref{eq:q}, 
    \begin{equation}\label{eq:qrelation}
        \begin{aligned}
            &\dot{q}_x(t)^2+\dot{q}_y(t)^2+2\mu\|\mathbf{q}(t)\|_2\frac{d\|\mathbf{q}(t)\|_2}{dt}+\mu^2\|\mathbf{q}(t)\|_2^2=u_i(t)^2.\\
        \end{aligned}
    \end{equation}
    We also have
    \begin{equation}\label{eq:dqdt}
    \begin{aligned}
        &\left(\frac{d\|\mathbf{q}(t)\|_2}{dt}\right)^2=\left(\frac{d\sqrt{q_x^2(t)+q_y^2(t)}}{dt}\right)^2\\
        &=\frac{q_x^2(t)\dot{q}_x^2(t)+2q_x(t)\dot{q}_x(t)q_y(t)\dot{q}_y(t)+q_y^2(t)\dot{q}_y^2(t)}{q_x^2(t)+q_y^2(t)}\\
        &=\dot{q}_x^2(t)+\dot{q}_y^2(t)\\
        &-\frac{q_y^2(t)\dot{q}_x^2(t)-2q_x(t)\dot{q}_x(t)q_y(t)\dot{q}_y(t)+q_x^2(t)\dot{q}_y^2(t)}{q_x^2(t)+q_y^2(t)}\\
        &=\dot{q}_x^2(t)+\dot{q}_y^2(t)-\frac{(q_y(t)\dot{q}_x(t)-q_x(t)\dot{q}_y(t))^2}{q_x^2(t)+q_y^2(t)}.
    \end{aligned}
    \end{equation}
    By Substituting \eqref{eq:dqdt} into \eqref{eq:qrelation}, we have:
    $$
    \frac{d\|\mathbf{q}(t)\|_2}{dt}+\mu\|\mathbf{q}(t)\|_2\leq u_i(t).
    $$
    
    According to Gronwall-Bellman Inequality, $\|\mathbf{q}(t)\|_2\leq \frac{u_{im}}{\mu}(1-e^{-\mu t})$. 
    Substituting this into \eqref{eq:xr} yields
    \begin{equation}
        \|\mathbf{x}_i(t)-\mathbf{x}_{ic}(t)\|_2\leq\int_0^t\frac{u_{im}}{\mu}(1-e^{-\mu \tau})d\tau=r_{ic}(t).
    \end{equation}
    Therefore, a player can only reach the points inside the isochron at $t$.  
    
    And for any point $\mathbf{x}$ inside an isochron at $t$, the player can reach it by reducing the amplitude of  acceleration to $\hat{u}_i=\frac{\mu}{t-\frac{1-e^{\mu t}}{\mu}}\|\mathbf{x}_{ic}(t)-\mathbf{x}\|_2$.
\endproof

\section{Proof of Theorem~\ref{thm:circumscribe}}\label{sec:ap-tangent}
\begin{proof}
We provide the proof for the externally tangent case; the internally tangent case can be treated analogously. Let $o(t)=\|\mathbf{x}_{Ac}(t)-\mathbf{x}_{Dc}(t)\|_2^2$, $p(t)=(r_{Ac}(t)+r_{Dc}(t))^2$, 
{\vspace{-0.8cm}
\footnotesize
\begin{equation*}
    \begin{aligned}
        o(t) = &\|\Delta\mathbf{x}\|_2^2+\frac{\|\Delta\mathbf{v}\|_2^2}{\mu^2}(1-e^{-\mu t})^2+2\frac{\Delta\mathbf{x}^T}{\mu}\Delta\mathbf{v}(1-e^{-\mu t}),\\
        p(t) = &\frac{(u_{Am}+u_{Dm})^2}{\mu^2}\left(t-\frac{1-e^{-\mu t}}{\mu}\right)^2,
    \end{aligned}
\end{equation*}}
where $\Delta\mathbf{x} = \mathbf{x}_{A0}-\mathbf{x}_{D0},\ \Delta\mathbf{v}=\mathbf{v}_{A0}-\mathbf{v}_{D0}$. The condition of external tangent is $o(t)-p(t)=0$.
Given $\|\Delta\mathbf{x}\|_2>0$, it is obvious that $o(0)-p(0)>0$ and $o(+\infty)-p(+\infty)<0$. We analyse the number of zeros points of $o-p$ by analysing their derivatives with respect to $t$.
{\vspace{-0.8cm}
\footnotesize
\begin{equation*}
    \begin{aligned}
        &o'=2e^{-\mu t}\left(\frac{\|\Delta\mathbf{v}\|_2^2}{\mu}(1-e^{-\mu t})+\Delta\mathbf{x}^T\Delta\mathbf{v}\right),\\
        &o''=2e^{-\mu t}(\|\Delta\mathbf{v}\|_2^2(2e^{-\mu t}-1)-\mu\Delta\mathbf{x}^T\Delta\mathbf{v}),\\
        &o^{(3)}=2\mu e^{-\mu t}(\mu\Delta\mathbf{x}^T\Delta\mathbf{v}-\|\Delta\mathbf{v}\|_2^2(4e^{-\mu t}-1)),\\
        &p'=2\frac{(u_{Am}+u_{Dm})^2}{\mu^2}\left(t-\frac{1-e^{-\mu t}}{\mu}\right)(1-e^{\mu t}),\\
        &p''=2\frac{(u_{Am}+u_{Dm})^2}{\mu^2}\left((1-e^{-\mu t})^2+\mu e^{-\mu t}\left(t-\frac{1-e^{-\mu t}}{\mu}\right)\right),\\
        &p^{(3)}=2\frac{(u_{Am}+u_{Dm})^2}{\mu^2}e^{-\mu t}(4(1-e^{-\mu t})-\mu t).
    \end{aligned}
\end{equation*}}
The situation can be divided into 3 categories based on the monotonicity of $o$, which depends on $\Delta\mathbf{x}^T\Delta\mathbf{v}$ and $\|\Delta\mathbf{v}\|_2^2+\mu\Delta\mathbf{x}^T\Delta\mathbf{v}$. 

\textbf{type 1:} $\|\Delta\mathbf{v}\|_2^2+\mu\Delta\mathbf{x}^T\Delta\mathbf{v}<0$. Since $o'<0,p'\geq0$, $o-p$ decreases monotonously and has only one zero point.

\textbf{type 2:} $\|\Delta\mathbf{v}\|_2^2+\mu\Delta\mathbf{x}^T\Delta\mathbf{v}\geq0,\ \Delta\mathbf{x}^T\Delta\mathbf{v}\leq0$. We first investigate the zero point number of $o''-p''$. 

$p''(t)>0,\ \forall t>0$ and $p''(0)=0$. $o''(t)>0$ for $t\in[0,t_M)$ where $t_M=\frac{1}{\mu}\ln{\frac{2\|\Delta\mathbf{v}\|_2^2}{\|\Delta\mathbf{v}\|_2^2+\mu\Delta\mathbf{x}^T\Delta\mathbf{v}}}$. $o''(t_M)=0$. $o''$ decreases monotonously when $t<t_M$. We only consider $t<t_M$ since only in this period could $o''-p''=0$.

If $o''-p''$ has more than one zero point, then the following statement holds:
$\exists t_1<t_2<t_M,\ \text{s.t.}\ o''(t_1)-p''(t_1)<0,\ o''(t_2)-p''(t_2)>0.$
According to Lagrange's mean value theorem, there exists a time $t_3$ such that $t_1<t_3<t_2$, $p^{(3)}(t_3)<o^{(3)}(t_3)<0$. The expression for the ratio of $p^{(3)}$ and $o^{(3)}$ is given by: $$\frac{p^{(3)}}{o^{(3)}}=\frac{\frac{(u_{Am}+u_{Dm})^2}{\mu^2}(4(1-e^{-\mu t})-\mu t)}{\mu (\|\Delta\mathbf{v}\|_2^2+\mu\Delta\mathbf{x}^T\Delta\mathbf{v}-4\|\Delta\mathbf{v}\|_2^2e^{-\mu t})}.$$
The denominator increases monotonously, while the numerator decreases monotonously when $p^{(3)}(t)<0$. Therefore, $p^{(3)}(t)<o^{(3)}(t),\forall t>t_3$. This implies $p''(t)<o''(t), \forall t>t_2$, which is contradict to the fact that $p''(t_M)>o''(t_M)$. Hence, there is only one zero point for $o''-p''$. 
This implies that $o'-p'$ initially increases from $2\Delta\mathbf{x}^T\Delta\mathbf{v}$ and then decreases to $-\infty$. There are at most two zeros points for $o'-p'$, denoted as $t_{o1}, t_{o2}$. So, the function $o-p$ decreases for $t<t_{o1}$, increases for $t_{o1}<t<t_{o2}$, and decreases again for $t>t_{o2}$, having at most three zero points.

\textbf{type 3:} $\Delta\mathbf{x}^T\Delta\mathbf{v}>0$.
We will first prove that $o'-p'$ has only one zero point. 
If $\|\Delta\mathbf{v}\|_2^2\leq\mu\Delta\mathbf{x}^T\Delta\mathbf{v}$, then $o''(t)<0,\forall t$, which means $o'-p'$ decreases monotonously from $2\Delta\mathbf{x}^T\Delta\mathbf{v}$ to $-\infty$. 
If $\|\Delta\mathbf{v}\|_2^2>\mu\Delta\mathbf{x}^T\Delta\mathbf{v}$, it is necessary to determine the number of zeros points of $o''-p''$. The value of $p^{(3)}$ at $t_M$ is given by:
\begin{equation*}
    p^{(3)}(t_M)=\frac{2(u_{Am}+u_{Dm})^2}{\mu^2e^{\mu t}}\left(2(1-k)+\ln\frac{1+k}{2}\right),
\end{equation*}
where $k=\frac{\mu\Delta\mathbf{x}^T\Delta\mathbf{v}}{\|\Delta\mathbf{v}\|_2^2}\in (0,1)$. Since $2(1-k)+\ln\frac{1+k}{2}>0$ for $k\in(0,1)$, $p^{(3)}(t_M)>0$. This implies that $p''$ increases for $t<t_M$. So, $o''-p''$ only has one zero point. 

The function $o'-p'$ initially increases from $2\Delta\mathbf{x}^T\Delta\mathbf{v}$, during which it does not intersect with $x=0$. It then decreases to $-\infty$. Hence, $o'-p'$ has one zero point.
Therefore, $o-p$ initially increases from $\|\Delta\mathbf{x}\|_2^2$ and then decreases to $-\infty$, hence has one zero point.
\end{proof}

\begin{corollary}\label{cor:reach_point}
    Any point $\mathbf{x}\in\mathbb{R}^2$ can be reached by a player via at least one and at most three Normal strategies.
\end{corollary}
\begin{proof}
    Consider the case where $u_{Am}=0$, $\mathbf{v}_{A0}=[0,0]^T$, and $\mathbf{x}_{A0}=\mathbf{x}$. In this case, $\mathcal{I}_A(t)$ degenerates to the point $\mathbf{x}$; thus, the condition that $\mathcal{I}_D(t)$ passes through $\mathbf{x}$ is equivalent to saying that $\mathcal{I}_D(t)$ is tangent to $\mathcal{I}_A(t)$. 
    By Theorem~\ref{thm:circumscribe}, $\mathcal{I}_D(t)$ intersects $\mathbf{x}$ at least once and at most three times. It follows that the defender has at least one and at most three Normal strategies to reach $\mathbf{x}$. The case for the attacker can be treated similarly.
\end{proof}

% \section{}
% The regions with $t_{D1}<t_A<t_{D2}$ can be further divided into 7 categories according to the relations between $t_{Ai}(\mathbf{x})$ and $t_{Dj}(\mathbf{x})$:
% \begin{enumerate}
%     \item $t_{D1}<t_{A1}<t_{D2}<t_{A2}<t_{D3}<t_{A3}$
%     \item $t_{D1}<t_{A1}<t_{D2}<t_{A2}<t_{A3}<t_{D3}$
%     \item $t_{D1}<t_{A1}<t_{D2}<t_{D3}<t_{A2}<t_{A3}$
%     \item $t_{D1}<t_{A1}<t_{A2}<t_{D2}<t_{A3}<t_{D3}$
%     \item $t_{D1}<t_{A1}<t_{A2}<t_{A3}<t_{D2}<t_{D3}$
%     \item $t_{D1}<t_{A1}<t_{A2}<t_{D2}<t_{D3}<t_{A3}$
%     \item $t_{D1}<t_{A}<t_{D2}<t_{D3}$
% \end{enumerate}

\bibliographystyle{unsrt}
\bibliography{./bibtex/bib/IEEEexample}

\end{document}